\documentclass[pre,a4paper,superscriptaddress,twocolumn,showpacs,amsmath,amssymb,floatfix]{revtex4}

\def\<{{<}}
\def\>{{>}}

\usepackage{graphicx}
\usepackage{amssymb}
\usepackage{float}

\input{epsf}

\begin{document}

\title{Google matrix of the citation network of Physical Review}

\author{Klaus M. Frahm}
\affiliation{\mbox{Laboratoire de Physique Th\'eorique du CNRS, IRSAMC, 
Universit\'e de Toulouse, UPS, 31062 Toulouse, France}}
\author{Young-Ho Eom}
\affiliation{\mbox{Laboratoire de Physique Th\'eorique du CNRS, IRSAMC, 
Universit\'e de Toulouse, UPS, 31062 Toulouse, France}}
\author{Dima L. Shepelyansky}
%\homepage[]{http://www.quantware.ups-tlse.fr}
\affiliation{\mbox{Laboratoire de Physique Th\'eorique du CNRS, IRSAMC, 
Universit\'e de Toulouse, UPS, 31062 Toulouse, France}}

\date{October 21, 2013}
%\date{\today}

\begin{abstract}
We study the statistical properties of spectrum and eigenstates of
the Google matrix of the citation network of Physical Review
for the period 1893 - 2009. 
The main fraction of  complex eigenvalues 
with largest modulus is determined numerically
by different methods based on high 
precision computations with up to $p=16384$ binary digits 
that allows to resolve hard numerical problems 
for small eigenvalues. The nearly nilpotent matrix structure allows
to obtain a semi-analytical computation of 
eigenvalues. We find that the spectrum is characterized by the fractal Weyl law
with a fractal dimension $d_f \approx 1$.
It is found that the majority of eigenvectors
are located in a localized phase. The statistical distribution
of articles in the PageRank-CheiRank plane is established
providing a better understanding of information flows on the network.
The concept of ImpactRank is proposed to determine
an influence domain of a given article.
We also discuss the properties of random matrix models 
of Perron-Frobenius operators.   
\end{abstract}

%89.20.Hh       World Wide Web, Internet
%89.75.Hc       Networks and genealogical trees 
%05.40.Fb       Random walks and Levy flights
%72.15.Rn Localization effects (Anderson or weak localization) 
%89.75.Fb Structures and organization in complex systems

\maketitle

\section{Introduction}
%%%%%%%%%%%%%%%%%%%%%%%%%%%%%%%%%%%%%%%%%%%%%%%%%%%%%%%%%
The development of Internet led to emergence of 
various types of complex directed networks 
created by modern society. The size of such networks
grows rapidly going beyond ten billions in last two decades
for the World Wide Web (WWW).  Thus the development of
mathematical tools for the statistical analysis 
of such networks becomes of primary importance. 
In 1998, Brin and Page 
proposed the analysis of WWW on the basis of PageRank
vector of the associated Google matrix constructed 
for a directed network   \cite{brin}.
The mathematical foundations of this analysis
are based on Markov chains \cite{markov}
and Perron-Frobenius operators \cite{mbrin}.
The PageRank algorithm allows to compute the ranking
of network nodes and is known to be at the heart 
of modern search engines \cite{googlebook}.
However, in many respects the statement of Brin and Page that
{\it ``Despite the importance of large-scale search engines on the web,
very little academic research has been done on them''} \cite{brin}
still remains valid at present. In our opinion, 
this is related to the fact that the Google matrix $G$
belongs to a new class of operators which 
had been rarely studied in physical systems.
Indeed, the physical systems are usually described by Hermitian
or unitary matrices for which the Random Matrix Theory \cite{mehta}
captures many universal properties. In contrast, the
Perron-Frobenium operators and Google matrix 
have eigenvalues distributed in the complex plane
belonging to another class of operators.

The Google matrix is constructed from 
the adjacency matrix $A_{ij}$ which has
unit elements if there is a  link pointing from node 
$j$ to node $i$ and zero otherwise.
Then the matrix of Markov transitions
is constructed by normalizing elements of each column to unity
($S_{ij}=A_{ij}/\sum_{i} A_{ij}$, $\sum_j S_{ij}=1$) 
and replacing columns with only zero elements 
({\em dangling nodes}) by $1/N$, with $N$ being the matrix size.
After that the Google matrix of the network takes the form
\cite{brin,googlebook}:
\begin{equation}
   G_{ij} = \alpha  S_{ij} + (1-\alpha)/N \;\; .
\label{eq1} 
\end{equation} 
The damping parameter $\alpha$ in the WWW context 
describes the probability 
$(1-\alpha)$ to jump to any node for a random surfer. 
For WWW the Google search engine uses 
$\alpha \approx 0.85$ \cite{googlebook}.
The PageRank vector $P_i$ is the right eigenvector 
of $G$ at $\lambda=1$ ($\alpha <1$).
According to the Perron-Frobenius theorem \cite{mbrin},
$P_i$ components are positive and represent the probability
to find a random surfer on a given node $i$ (in the stationary limit) 
\cite{googlebook}.
All nodes can be ordered in a decreasing 
order of probability $P(K_i)$ with 
highest probability at top values of PageRank
index $K_i=1,2,....$.

The distribution of eigenvalues of $G$ can be rather 
nontrivial with appearance of the fractal Weyl law
and other unusual properties (see e.g. \cite{zhirovulam,ermannweyl}).
For example, a matrix $G$ with  random positive matrix elements,
normalized to unity in each column,
has $N-1$ eigenvalues $\lambda$ concentrated in a 
small radius $|\lambda| < 1/\sqrt{3N}$ and one eigenvalue
$\lambda=1$ (see below in section \ref{sec_RPFM}).
Such a distribution is drastically different 
from the eigenvalue distributions found for
directed networks with algebraic distribution of links \cite{ggs1}
or those found numerically for other directed networks
including WWW of universities \cite{ggs2,univuk}, 
Linux Kernel and Twitter networks \cite{linux,twitter},  Wikipedia 
networks \cite{wikileo,wikievol}. In fact even the Albert-Barab\'asi model
of preferential attachment \cite{barabasi}
still generates the complex spectrum of $\lambda$
with a large gap ($|\lambda|< 1/2$) \cite{ggs1} being very different
from the  gapless and strongly degenerate
$G$ spectrum of WWW of British universities \cite{univuk}
and Wikipedia \cite{wikileo,wikievol}. Thus it is useful to get
a deeper understanding of the spectral properties of directed
networks and to develop more advanced models
of complex networks which have a spectrum similar
to such networks as British universities and Wikipedia.

With the aim to understand the spectral properties of 
Google matrix of directed networks we study here the Citation Network of
Physical Review (CNPR) for the whole period up to 2009 \cite{physrev_data}.
This network has $N=463348$ nodes (articles) and $N_\ell=4691015$
links. Its network structure is very similar to the tree
network since the citations are time ordered (with only a few exceptions
of mutual citations of simultaneously published articles).
As a result we succeed to develop powerful tools 
which allowed us to obtain the spectrum of $G$ in semi-analytical
way. These results are compared with the spectrum
obtained numerically with the help of the powerful Arnoldi method
(see its description in \cite{stewart,frahm2010}). 
Thus we are able to get a better understanding of 
the spectral properties of this network. Due to time ordering of 
article citations
there are strong similarities between the CNPR and the network of integers
studied recently in \cite{integers}.

We note that the PageRank analysis of the CNPR had been
performed in \cite{redner,maslov},\cite{fortunato,fortunato2}
showing its efficiency in determining the influential articles
of Physical Review. The citation networks 
are rather generic (see e.g. \cite{bergstrom}) 
and hence the extension of  PageRank analysis
of such networks is an interesting and important task.
Here we put the main accent on the spectrum and eigenstates
properties of the Google matrix of the CNPR
but we also discuss the properties of two-dimensional (2D) ranking on
PageRank-CheiRank plane developed 
recently in \cite{alik,zzswiki},\cite{2dmotor}. 
We also analyze the properties of 
%ImpactRank which shows how a given article is influenced by
%other articles and how it affects articles published after it.
ImpactRank which shows a domain of influence of a given article.

In addition to the whole CNPR we also consider
the CNPR without Rev. Mod. Phys. articles which
has $N=460422$,  $N_\ell=4497707$.
If in the  whole CNPR we eliminate future citations
(see description below)
then this triangular CNPR has
$N=463348$, $N_\ell=4684496$.
Thus on average we have approximately 10 links per node.
The network includes all articles
of Physical Review from its foundation in 1893
till the end of 2009.

The paper is composed as follows: in Section II we present a detailed analysis
of the Google matrix spectrum of CNPR, the fractal Weyl law
is discussed in Section III, properties of eigenstates
are discussed in Section IV, CheiRank versus PageRank distributions
are considered in Section V, properties of
impact propagation through the network are studied in 
Section VI, certain random matrix models of Google matrix are 
studied in Section VII,
the discussion of the results is given in Section VIII.

%\vskip -0.2cm
\section{Eigenvalue spectrum}

\begin{figure}[H]
\begin{center}
\includegraphics[width=0.48\textwidth]{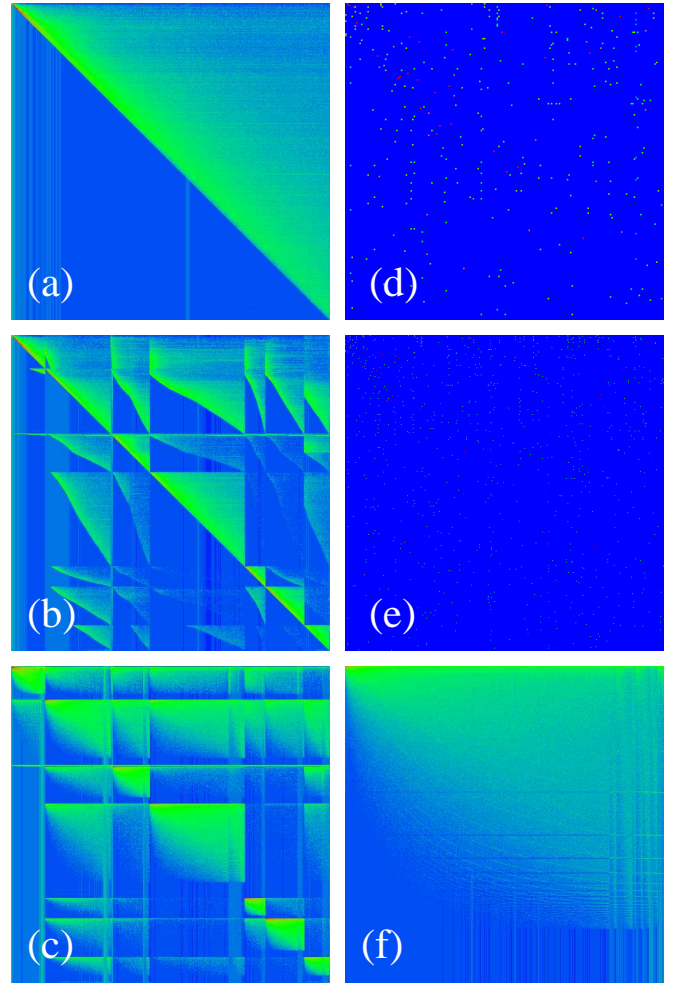}
\end{center}
\vglue -0.3cm
\caption{(Color online)
Different order representations of the Google matrix of the CNPR 
($\alpha=1$). 
The  left panels show: $(a)$ the density of matrix elements 
$G_{tt'}$ in the basis of the publication time index $t$ (and $t'$);
$(b)$ the density of matrix elements in the basis 
of journal ordering according to: Phys.~Rev.~Series~I, Phys.~Rev., 
Phys.~Rev.~Lett., Rev.~Mod.~Phys., Phys.~Rev.~A, B, C, D, E, Phys.~Rev.~STAB 
and Phys.~Rev.~STPER  with time ordering inside each journal;
$(c)$ the same as $(b)$ 
but with PageRank index ordering inside each journal. 
Note that the journals Phys.~Rev.~Series~I, Phys.~Rev.~STAB and 
Phys.~Rev.~STPER are not clearly visible due to a small number of published 
papers. Also Rev.~Mod.~Phys. appears only as a thin line with 2-3 pixels 
(out of 500) due to a limited number of published papers.  
The  panels $(a), (b), (c), (f) $ 
show the coarse-grained density of matrix elements 
done on $500 \times 500$ square cells for the entire network.
In panels $(d), (e), (f)$ the  matrix elements 
$G_{KK'}$ are shown in the basis of PageRank index $K$ (and $K'$) with 
the range 
$1 \leq K,K' \leq 200$ $(d)$;
$1 \leq K,K' \leq 400$ $(e)$;
$1 \leq K,K' \leq N$ $(f)$.
Color shows the amplitude (or density) of matrix elements $G$ 
changing from blue/black for zero value to red/grey at maximum value. 
The PageRank index $K$ 
is determined from the PageRank vector at $\alpha=0.85$. 
\label{fig1}}
\end{figure}

The Google matrix of CNPR is constructed 
on the basis of Eq.(\ref{eq1}) using citation links
from one article to another (see also \cite{maslov,fortunato,fortunato2}).
The matrix structure for different  order representations
of articles  is shown in Fig.~\ref{fig1}.
In the top left panel all articles are ordered by time 
that generates almost perfect triangular structure
corresponding to time ordering of citations.
Still there are a few cases with joint citations
of articles which appear almost at the same time.
Also there are dangling nodes which generate transitions
to all articles with elements $1/N$ in $G$.
This breaks the triangular structure and as we will see later 
it is just the combination of these dangling node contributions with 
the other non-vanishing matrix elements (see also 
below Eq. \ref{eq_matrixS}) which will allow to formulate 
a semi-analytical theory to determine the eigenvalue spectrum. 
%%%%%%%%%%%%%%%%%%%%%%%%%%%%%%%%%%%%%%%%%%%%%%%%%%%%%%%%%
%%%%%%%%%%%%%%%%%%%%%%%%%%%%%%%%%%%%%%%%%%%%%%%%%%%%%%%%%
\noindent
The triangular matrix 
structure is also well visible in the middle  left 
panel where articles are time ordered within 
each  Phys. Rev. journal. The left
bottom panel
shows the matrix elements for each Phys. Rev. journal
when inside each journal the articles are ordered by their PageRank index
$K$. The right panels show the matrix elements of $G$
on different scales,
when all articles are ordered by the PageRank index $K$. 
Top two right panels have a relatively small number of nonzero
matrix elements showing that top PageRank articles 
rarely quote other top PageRank articles.

\begin{figure}[H]
\begin{center}
\includegraphics[width=0.48\textwidth]{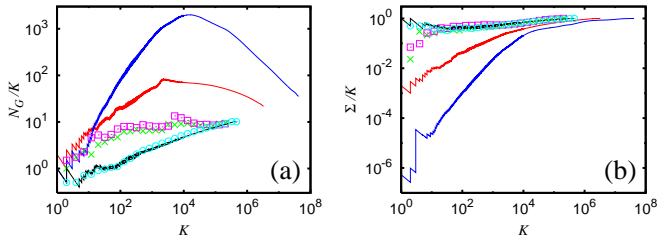}
\end{center}
\vglue -0.2cm
\caption{(Color online)
$(a)$ Dependence of the linear density $N_G/K$ 
of nonzero elements of 
the adjacency matrix among top PageRank nodes on the PageRank index $K$
for the networks of Twitter (top blue/black curve), 
Wikipedia (second from top red/gray curve), 
Oxford University 2006 (magenta/grey boxes), Cambridge University 2006 
(green/grey crosses), with data taken from Ref. \cite{twitter}, 
and Physical Review all journals (cyan/grey circles) and Physical Review 
without Rep.~Mod.~Phys. (bottom black curve). 
$(b)$ Dependence of the quantity $\Sigma/K$ 
on the PageRank index $K$ with $\Sigma=\sum_{K_1<K,\,K_2<K} G_{K_1,K_2}$
being the weight of the Google matrix elements inside the $K\times K$ 
square of top PageRank indexes. 
The curves correspond to the same networks as in $(a)$:
Physical Review without Rep.~Mod.~Phys. (top black curve), Physical Review 
all journals (cyan/grey circles), Oxford University 2006 
(magenta/grey boxes), 
Cambridge University 2006 (green/grey crosses), 
Wikipedia (second bottom red/grey curve), 
and Twitter (bottom blue/black curve). 
\label{fig2}}
\end{figure}

The dependence of number of no-zero links 
$N_G$, between nodes with PageRank index
being less than $K$, on $K$ is shown in Fig.~\ref{fig2}
(left panel).
We see that compared to the other networks 
of universities, Wikipedia and Twitter
studied in \cite{wikileo} we have for CNPR the lowest
values of $N_G/K$ practically for all available $K$ values.
This reflects weak links between top PageRank articles
of CNPR being in contrast with Twitter
which has very high interconnection between
top PageRank nodes. Since the matrix 
elements $G_{KK'}$ are inversely proportional to the number
of links we have very strong average
matrix elements for CNPR at top $K$
values (see Fig.~\ref{fig2} (right panel)).

In the following we present the
results of numerical and analytical analysis of the spectrum of 
the CNPR matrix $G$.

\subsection{Nearly nilpotent matrix structure} 

The triangular structure of the CNPR Google matrix 
in time index (see Fig. \ref{fig1}) has important consequences for the 
eigenvalue spectrum $\lambda$
defined by the equation for the eigenstates $\psi_i(j)$:
\begin{equation}
 \sum_{j'}  G_{jj'} \psi_i(j') = \lambda_i \psi_i(j) \; .
\label{eq2}  
\end{equation} 

\begin{figure}[H]
\begin{center}
\includegraphics[width=0.48\textwidth]{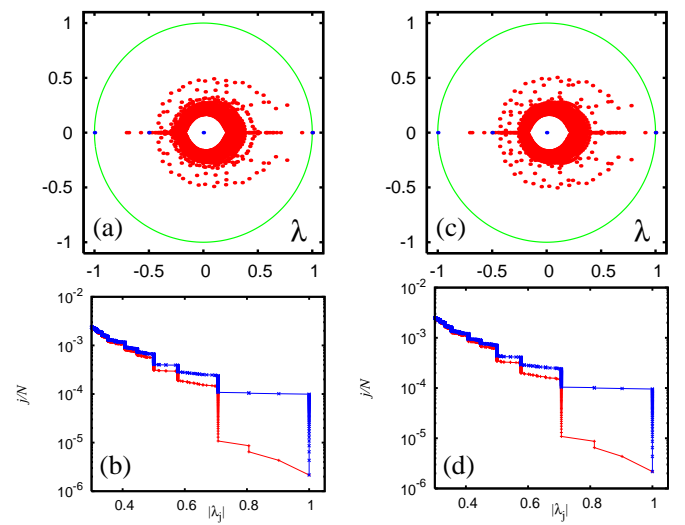}
\end{center}
\vglue -0.3cm
\caption{(Color online)
Spectrum of $S$  for CNPR (reduced CNPR without 
Rev.~Mod.~Phys.) shown on left panels (right panels). 
{\em Panels (a), (c):} Subspace eigenvalues (blue/black dots)  and 
core space eigenvalues (red/grey dots) in $\lambda$-plane 
(green/grey curve shows unit circle);
there are 27 (26) invariant subspaces, with maximal dimension 6 (6) 
and the sum of all subspace dimensions is $N_s=71$ (75). 
The core space eigenvalues are obtained from the Arnoldi method applied to 
the core space subblock $S_{cc}$ of $S$ with Arnoldi dimension $n_A=8000$ as 
explained in Ref.~\cite{univuk} and using standard double-precision 
arithmetic.
{\em Panels (b), (d):} 
Fraction $j/N$ of eigenvalues, shown in a logarithmic scale, 
with $|\lambda| >  |\lambda_j|$ for the 
core space eigenvalues 
(red/grey bottom curve) and all eigenvalues (blue/black top curve)
from raw data of top panels. The number of 
eigenvalues with $|\lambda_j|=1$ is 45 (43) of which 
27 (26) are at $\lambda_j=1$; this number is identical to the 
number of invariant subspaces which have each one unit eigenvalue.
\label{fig3}}
\end{figure}

The spectrum of $G$ at $\alpha=1$, or the spectrum of $S$,
obtained by the Arnoldi method \cite{stewart,frahm2010}
with the Arnoldi dimension   $n_A=8000$,
is shown in Fig.~\ref{fig3}. For comparison we also show the
case of reduced CNPR without Rev. Mod. Phys..
We see that the spectrum of the reduced case 
is rather similar to the spectrum of the full CNPR.

The nodes can be decomposed in invariant subspace nodes and 
core space nodes and the matrix $S$ can be written in the block structure
\cite{univuk}: 
\begin{equation}
\label{eq3}
S=\left(\begin{array}{cc}
S_{ss} & S_{sc}  \\
0 & S_{cc}  \\
\end{array}\right)
\end{equation}
where $S_{ss}$ contains the links from subspace nodes to other subspace nodes, 
$S_{cc}$ the links from core space nodes to core space nodes and 
$S_{sc}$ some coupling links from the core space to the invariant subspaces. 
The subspace-subspace block $S_{ss}$ is actually composed of (potentially) 
many diagonal blocks for each of the invariant subspaces. Each of these blocks 
corresponds to a column sum normalized matrix of the same type as $G$ 
and has therefore at least one unit eigenvalue thus explaining 
a possible high degeneracy of the eigenvalue $\lambda=1$ of $S$. 
This structure is discussed in detail in \cite{univuk}. The university 
networks discussed in \cite{univuk} had a considerable number of subspace 
nodes (about 20~\%) with a high degeneracy $\sim 10^3$ of the leading 
unit eigenvalue. However, for the CNPR the number of subspace nodes and unit 
eigenvalues is quite small (see figure caption of Fig.~\ref{fig3} for 
detailed values). 

A network with a similar triangular structure, 
constructed from factor decompositions of integer numbers, 
was previously studied in \cite{integers}. There it was analytically 
shown that the 
corresponding matrix $S$ has only a small number of non-vanishing 
eigenvalues and that the numerical 
diagonalization methods, including the Arnoldi method, are facing subtle 
difficulties of numerical stability due to large Jordan blocks associated 
to the highly degenerate zero eigenvalue. The numerical diagonalization 
of these Jordan blocks is highly sensitive to numerical round-off errors. For 
example a perturbed Jordan block of dimension $D$ associated to the eigenvalue 
zero and with a perturbation $\varepsilon$ in the opposite corner has 
eigenvalues on a complex circle of radius $\varepsilon^{1/D}$ \cite{integers}
which may became quite large for sufficient large $D$ even for 
$\varepsilon\sim 10^{-15}$. Therefore in presence of many such Jordan blocks 
the numerical diagonalization methods 
create rather big ``artificial clouds'' 
of incorrect eigenvalues. 

In the examples studied in \cite{integers} these 
clouds extended up to eigenvalues $|\lambda|\approx 0.01$. 
The spectrum for the Physical Review network shown in Fig. \ref{fig3} shows 
also a sudden increase of the density of eigenvalues below 
$|\lambda|\approx 0.3-0.4$ and one needs 
to be concerned if these eigenvalues 
are numerically correct or only an artifact of the same type of numerical 
instability. 
Actually, there is a quite simple way to verify that they are not 
reliable due to problems in the numerical evaluation. 
For this we apply to the network or 
the numerical algorithm (in the computer program) certain transformations or 
modifications which are {\em mathematically neutral} 
or {\em equivalent}, e.~g. a permutation of the index numbers of 
the network nodes but keeping the same network-link structure, or simply 
changing the evaluation order in the sums used for the scalar products 
between vectors (in the Gram-Schmidt orthogonalization for the Arnoldi 
method). 
All these modifications should in theory not modify the results 
(assuming that all computations could be done with {\em infinite precision}) 
but in numerical computations on a computer with finite precision 
they modify the round-off errors. It turns indeed out that 
the modifications of the initially small round-off errors induce 
very strong, completely random modifications, for all eigenvalues 
below $|\lambda|\approx 0.3-0.4$ clearly indicating that the latter 
are numerically not accurate. Apparently the problematic 
numerical eigenvalue 
errors due to large Jordan blocks $\sim \varepsilon^{1/D}$ with 
$D\sim 10^2$ is quite stronger in the Physical Review 
citation network than in 
the previsously studied integer network \cite{integers}.

%%%%%%%%%%%%%%%%%%%%%%%%%%%%%%%%%%%%%%%%%%%%%%%%%%%%%%%%%
%%%%%%%%%%%%%%%%%%%%%%%%%%%%%%%%%%%%%%%%%%%%%%%%%%%%%%%%%

The theory of \cite{integers} is based on the exact triangular structure 
of the matrix $S_0$ 
which appears in the representation of $S=S_0+e d^T/N$ 
(see also below Eq. \ref{eq_matrixS}). 
In fact the matrix $S_0$ is obtained from the adjacency matrix by 
normalizing the sum of the elements in non-vanishing columns to unity and 
simply keeping at zero vanishing columns. For the network of integers
\cite{integers} this matrix is nilpotent with 
$S_0^l=0$ for a certain modest value of $l$ being much smaller than the 
network size $l\ll N$.
The nilpotency is very relevant in the paper for two reasons: first
it is responsible for the numerical problems to compute the
eigenvalues by standard methods (see next point) and second it is also
partly the solution by allowing a semi-analytical approach to
determine the eigenvalues in a different way.
 
For CNPR   the matrix $S_0$ is 
not exactly nilpotent despite the overall triangular matrix structure 
visible in Fig. \ref{fig1}. Even though most of 
the non-vanishing matrix elements 
$(S_0)_{tt'}$ (whose total number is equal to the number 
of links $N_\ell=4691015$)
are in the upper triangle $t<t'$ there  are a few non-vanishing elements 
in the lower triangle $t > t'$
%or on the diagonal $t\ge t'$ 
(whose number is $12126$ corresponding to $0.26$~\% of the total number 
of links \cite{timeorderdiscuss}). 
The reason is that in most cases papers cite other papers published earlier 
but in certain situations for papers with 
close publication date the citation 
order does not always coincide with the publication order. In some cases 
two papers even mutually cite each other. In the following we will call 
these cases ``future citations''. The rare non-vanishing matrix 
elements due to future citations are not visible 
in the coarse grained matrix representation of 
Fig. \ref{fig1} but they are responsible for the fact that $S_0$ of CMPR 
is not nilpotent and that there are also a few 
invariant subspaces. On a purely triangular network one can 
easily show the absence of invariant subspaces 
(smaller than the full network 
size) when taking into account the extra columns due to the dangling nodes. 

However, despite the effect of the future citations the matrix $S_0$ 
is still partly nilpotent. This can be seen by multiplying
a uniform initial vector $e$ (with all components being 1) 
by the matrix $S_0$ 
and counting after each iteration the number $N_i$ of non-vanishing entries 
\cite{nonzeroentry} 
in the resulting vector $S_0^i e$. For a nilpotent matrix $S_0$ 
with $S_0^l=0$ the number $N_i$ becomes obviously zero for $i\ge l$. 
On the other hand, since the components of $e$ and the non-vanishing matrix 
elements of $S_0$ are positive, one can easily verify that the condition 
$S_0^l e=0$ for some value $l$ also implies $S_0^l\psi=0$ for an arbitrary 
initial (even complex) vector $\psi$ which shows that $S_0$ must be nilpotent 
with $S_0^l=0$. 
%%%%%%%%%%%%%%%%%%%%%%%%%%%%%%%%%%%%%%%%%%%%%%%%%%%%%%%%%
%%%%%%%%%%%%%%%%%%%%%%%%%%%%%%%%%%%%%%%%%%%%%%%%%%%%%%%%%

\begin{figure}[H]
\begin{center}
\includegraphics[width=0.48\textwidth]{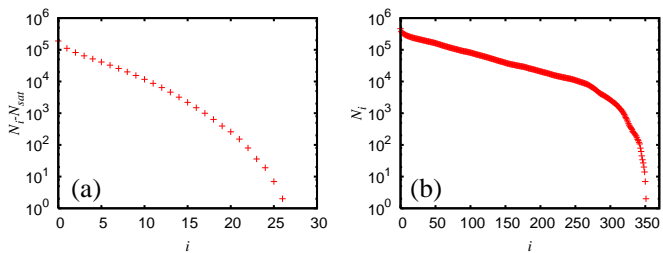}
\end{center}
\vglue -0.2cm
\caption{(Color online)
Number of occupied nodes $N_i$ (i.e. positive elements) 
in the vector $S_0^i\,e$ versus iteration number $i$ (red/grey crosses) 
for the  CNPR $(a)$ 
and the triangular  CNPR $(b)$. 
In both cases the initial value is the network size $N_0=N=463348$. 
For the CNPR $N_i$ saturates at 
$N_{i}=N_{sat}=273490\approx 0.590 N$ for $i\ge 27$ while for the 
triangular CNPR $N_i$ saturates at 
$N_{i}=0$ for $i\ge 352$ confirming the nilpotent structure of $S_0$. 
In  panel $(a)$ the quantity $N_i-N_{sat}$ is shown in order to 
increase visibility in the logarithmic scale.
\label{fig4}}
\end{figure}

In Fig.~\ref{fig4} we see that for the  CNPR the value 
of $N_i$ saturates at a value $N_{sat}=273490$ for $i\ge 27$ which is $59$~\% 
of the total number of nodes $N=463348$ in the network. On one hand the 
(small) number of future citations ensures 
that the saturation value of $N_i$ 
is not zero but on the other hand it is smaller than the total number of 
nodes by a macroscopic factor. Mathematically the first iteration 
$e\to S_0 e$ removes the nodes corresponding 
to empty (vanishing) lines of the 
matrix $S_0$ and the next iterations remove the nodes whose lines in $S_0$ 
have become empty after having removed from the network the non-occupied 
nodes due to previous iterations. For each node removed during this 
iteration process one can construct a vector belonging 
to the Jordan subspace 
of $S_0$ associated to the eigenvalue 0. In the following we call this 
subspace {\em generalized kernel}. It contains all eigenvectors of $S_0^j$ 
associated to the eigenvalue 0 where the integer $j$ is the 
size of the largest 
$0$-eigenvalue Jordan block. Obviously the dimension of this generalized 
kernel of $S_0$ is larger or equal than $N-N_{sat}=189857$ but we will 
see later that its actual dimension is even larger and quite 
close to $N$. We will argue below that most (but not all) 
of the vectors in the generalized kernel 
of $S_0$ also belong to the generalized kernel of $S$ which differs from $S_0$ 
by the extra contributions due to the dangling nodes. 
The high dimension of the generalized kernel containing many 
large $0$-eigenvalue Jordan subspaces explains 
very clearly the numerical problem due to which 
the eigenvalues obtained by the double-precision Arnoldi method 
are not reliable for $|\lambda|<0.3-0.4$.
%%%%%%%%%%%%%%%%%%%%%%%%%%%%%%%%%%%%%%%%%%%%%%%%%%%%%%%%%
%%%%%%%%%%%%%%%%%%%%%%%%%%%%%%%%%%%%%%%%%%%%%%%%%%%%%%%%%

\subsection{Spectrum for the triangular CNPR}

In order to extend  the theory  for 
the triangular matrices developed in  \cite{integers}
we consider the triangular CNPR obtained 
by removing all future citation links $t'\to t$ 
with $t\ge t'$ from the original CNPR. 
The resulting matrix $S_0$ of this reduced network is now indeed 
nilpotent with $S_0^{l-1}\neq 0$, $S_0^l=0$ and $l=352$ 
which is much smaller than the network size. This is clearly seen 
from Fig.~\ref{fig4} showing that $N_i$, calculated from 
the triangular CNPR,  indeed saturates 
at $N_i=0$ for $i\ge 352$. According to the 
arguments of \cite{integers}, and additional demonstrations 
given below, there are at most only 
$l=352$ non-zero eigenvalues of the Google matrix at $\alpha=1$. 
This matrix has the form
\begin{equation}
\label{eq_matrixS}
S=S_0+(1/N)\,e\,d^T
\end{equation}
where $d$ and $e$ are two vectors with $e(n)=1$ for all nodes 
$n=1,\,\ldots,\,N$ 
and $d(n)=1$ for dangling nodes $n$ (corresponding to vanishing columns 
in $S_0$) and $d(n)=0$ for the other nodes. In the following we  
call $d$ the dangling vector. The extra contribution 
$e\,d^T/N$ just replaces the empty columns (of $S_0$) with 
$1/N$ entries at each element and $d^T$ is the line vector obtained 
as the transpose of the column vector $d$. In Appendix A we 
extend the approach of  \cite{integers} showing analytically that 
the matrix $S$ has exactly $l=352\ll N$ non-vanishing eigenvalues 
which are given as the zeros of the reduced polynomial given in Eq. 
(\ref{eq_polyred}) and that it is possible to define a closed representation 
space for the matrix $S$ of dimension $l$ leading 
to an $l\times l$ representation matrix $\bar S$ given by Eq. 
(\ref{eq_repmatrix}) whose eigenvalues are exactly the zeros of 
the reduced polynomial. 

%%%%%%%%%%%%%%%%%%%%%%%%%%%%%%%%%%%%%%%%%%%%%%%%%%%%%%%%%
\begin{figure}[H]
\begin{center}
\includegraphics[width=0.48\textwidth]{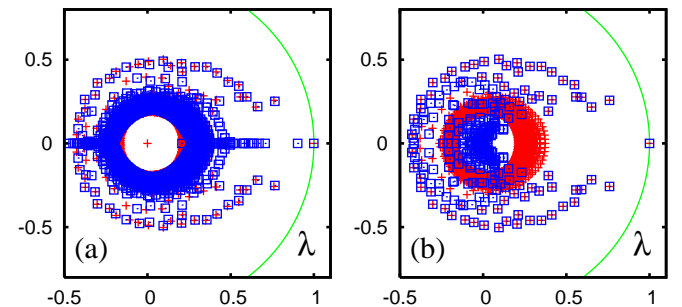}
\end{center}
\caption{(Color online)
{\em Panel (a):} Comparison of the core space eigenvalue spectrum 
of  $S$ for CNPR (blue/black squares) and 
triangular CNPR (red/grey crosses). Both 
spectra are calculated by the Arnoldi method with $n_A=4000$ and 
standard double-precision.
{\em Panel (b):} Comparison of the numerically determined 
non-vanishing 352 eigenvalues obtained from the 
representation matrix (\ref{eq_repmatrix}) (blue/black squares) with 
the spectrum of triangular CNPR (red/grey crosses) already 
shown in the left panel. 
Numerics is done with standard double-precision.
\label{fig5}}
\end{figure}
%%%%%%%%%%%%%%%%%%%%%%%%%%%%%%%%%%%%%%%%%%%%%%%%%%%%%%%%%

In the left panel of Fig.~\ref{fig5} we compare the core space
spectrum of $S$ for CNPR and triangular CNPR
(data are obtained by the Arnoldi method 
with  $n_A=4000$ and  standard double-precision).  We 
see that the largest complex eigenvalues 
are rather close for both cases 
but in the full network we have a lot of eigenvalues 
on the real axis (with $\lambda<-0.3$ or $\lambda>0.4$) which are absent 
for the triangular CNPR. Furthermore, both cases suffer from the 
same problem of numerical instability due to large Jordan blocks.

In the right panel of Fig. \ref{fig5} we compare the 
numerical double-precision spectra of the 
representation matrix ${\bar S}$ with the results of the 
Arnoldi method with double-precision and the uniform 
initial vector $e$ as start vector for the Arnoldi iterations 
(applied to the triangular CNPR). 
In Appendix B we explain that the Arnoldi method with this 
initial vector should in theory (in absence of rounding errors) 
also exactly provide the $l$ eigenvalues of ${\bar S}$ since by construction 
it explores the same $l$ dimensional $S$-invariant representation space 
that was used for the construction of ${\bar S}$ (in Appendix A). The fact 
that both spectra of the right panel of Fig. \ref{fig5} differ is therefore 
a clear effect of numerical errors and actually both cases 
suffer from different numerical problems (see Appendix B for details). 
A different, and in principle highly efficent, computational method is to 
calculate 
the spectrum of the triangular CNPR by determining numerically the $l$ zeros 
of the reduced polynomial (\ref{eq_polyred}) but according to the 
further discussion in Appendix B there are also numerical problems for this. 
Actually this method requires the help the 
GNU Multiple Precision Arithmetic Library (GMP library) 
\cite{gmplib} using 256 binary digits. Also the Arnoldi method can be 
improved by GMP library (see Appendix B for details) 
even though this is quite expensive in computational 
time and memory usage but still feasible (using up to $1280$ binary digits). 
Below we will also present results (for the spectrum of the full CNPR) 
based on a new method using the GMP library with up to $16384$ binary digits. 

\begin{figure}[H]
\begin{center}
\includegraphics[width=0.48\textwidth]{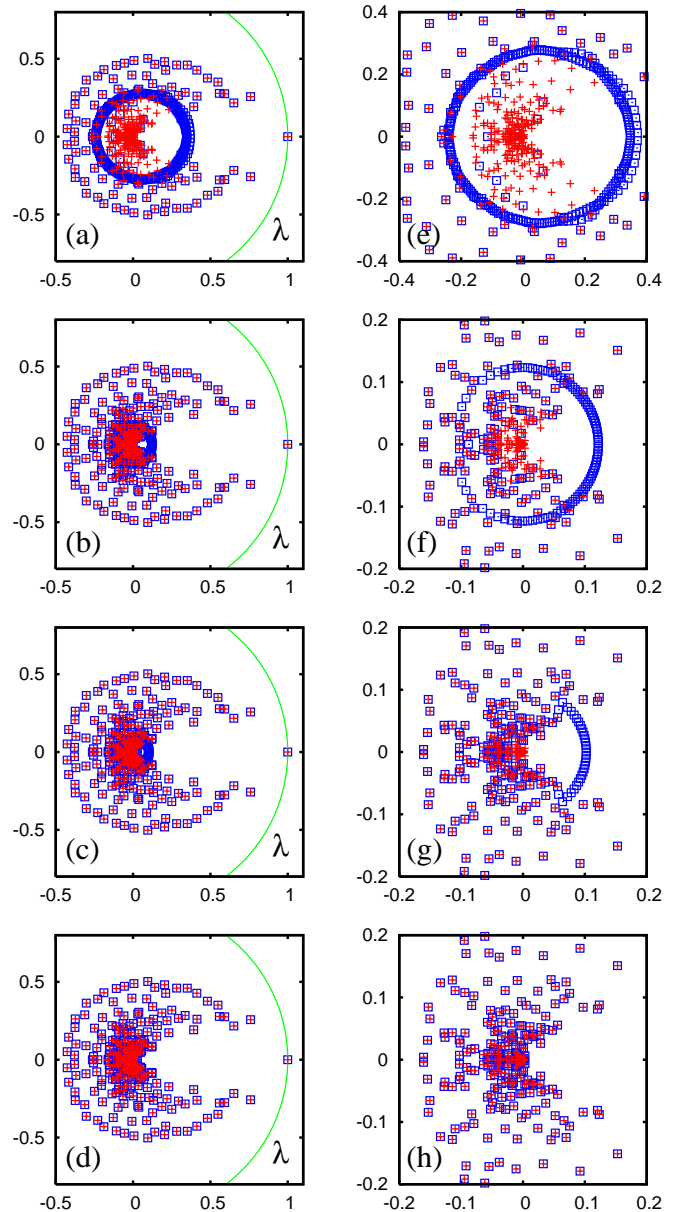}
\end{center}
\caption{(Color online)
Comparison of the numerically accurate 352 non-vanishing 
eigenvalues of $S$ matrix of triangular CNPR, determined 
by the Newton-Maehly method applied to the reduced polynomial 
(\ref{eq_polyred}) with a high-precision calculation of 
256 binary digits (red/grey crosses, all panels), 
with eigenvalues obtained by the Arnoldi method at different 
numerical precisions (for the determination of the Arnoldi matrix) 
for triangular CNPR and Arnoldi 
dimension $n_A=352$ (blue/black squares, all panels). 
The first row corresponds to the numerical precision of 52 binary digits 
for standard double-precision arithmetic. The second (third, fourth) row 
corresponds to the precision of 256 (512, 1280) binary digits.
All high precision calculations are done with the library 
GMP \cite{gmplib}. The panels in the left column show the complete spectra 
and the panels in the right columns show the spectra in a zoomed range: 
$-0.4\le \mbox{Re}(\lambda), \mbox{Im}(\lambda)\le <0.4$ for the first row 
or $-0.2\le \mbox{Re}(\lambda), \mbox{Im}(\lambda)\le 0.2$ for the 
second, third and fourth rows. 
\label{fig6}}
\end{figure}

In Fig.~\ref{fig6} we compare the exact spectrum 
of the triangular CNPR obtained by the 
high precision determination of the zeros of the reduced polynomial 
(using 256 binary digits) 
with the spectra of the Arnold method for 52 binary digits 
(corresponding to the 
mantissa of double-precision numbers), 
256, 512 and 1280 binary digits. Here we  
use for the Arnoldi method a uniform initial vector and the 
Arnold dimension $n_A=l=352$. In this case, as explained in Appendix B, 
in theory the Arnoldi method should provide the exact $l=352$ 
non-vanishing eigenvalues (in absence of round-off errors). 

However, with the precision of 52 bits we have a considerable number 
of eigenvalues on a circle of radius $\approx 0.3$ centered at $0.05$ 
indicating a strong influence of round-off errors due to the Jordan blocks. 
Increasing the precision to 256 (or 512) binary digits implies that the number of 
correct eigenvalue increases and the radius of this circle 
decreases to $0.13$ (or $0.1$) and in particular it does not 
extend to all angles.
We have to increase the precision of the Arnoldi method to 1280 binary digits 
to have a perfect numerical confirmation that the Arnoldi method explores 
the exact invariant subspace of dimension $l=352$ and generated by 
the vectors $v_j$ (see Appendix A). 
In this case the eigenvalues obtained from the Arnoldi method and the 
high-precision zeros of the reduced polynomial coincide with an error 
below $10^{-14}$ and in particular the Arnoldi method provides a nearly 
vanishing 
coupling matrix element at the last iteration confirming that there is indeed 
an exact decoupling of the Arnoldi matrix and an invariant closed subspace 
of dimension 352.

%%%%%%%%%%%%%%%%%%%%%%%%%%%%%%%%%%%%%%%%%%%%%%%%%%%%%%%%%
%%%%%%%%%%%%%%%%%%%%%%%%%%%%%%%%%%%%%%%%%%%%%%%%%%%%%%%%%

The results shown in Fig.\ref{fig6} clearly confirm the above theory 
and the scenario of the strong influence of Jordan blocks on the round-off 
errors. In particular, we find that in order to increase the 
numerical precision it is 
only necessary to implement the first step of the method, the Arnoldi 
iteration, using high precision numbers while the numerical diagonalization 
of the Arnoldi representation matrix can still be done using standard 
double-precision arithmetic. We also observe, that even for the 
case with lowest precision of 52 binary digits the eigenvalues obtained by the 
Arnoldi method are numerically accurate provided that there are well 
outside the circle (or cloud) of numerically incorrect eigenvalues. 

\subsection{High precision spectrum of the whole CNPR} 

\begin{figure}[H]
\begin{center}
\includegraphics[width=0.48\textwidth]{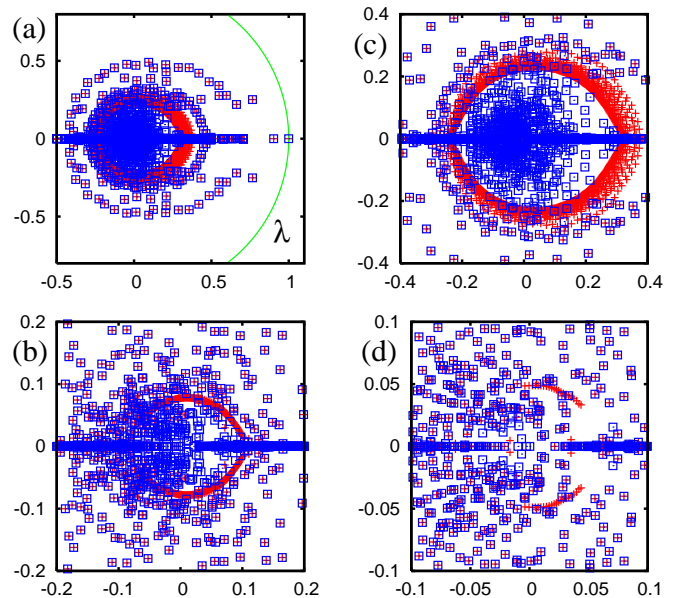}
\end{center}
\caption{ (Color online)
Comparison of the core space eigenvalue spectrum of $S$ of CNPR,
obtained by the high precision Arnoldi method 
using 768 binary digits (blue/black squares, all panels), 
with lower precision data of the Arnoldi method (red/grey crosses). 
In both top panels the red/grey crosses correspond to double-precision 
with 52 binary digits (extended range in $(a)$  and zoomed 
range in $(c)$). In the bottom $(b)$ (or $(d))$) panel red/grey crosses 
correspond to the numerical precision of 256 (or 512) binary digits. In these 
two cases only a zoomed range is shown. The eigenvalues outside the zoomed 
ranges coincide for both data sets up to graphical precision. 
In all cases the Arnoldi dimension is $n_A=2000$. 
High precision calculations are done with the library GMP \cite{gmplib}. 
\label{fig7}}
\end{figure}

Based on the observation that a high precision implementation of the 
Arnoldi method is useful for the triangular CNPR, 
we now apply the high precision Arnoldi method with 256, 512 and 756 
binary digits and $n_A=2000$ to the original CNPR. 
The results for the core space eigenvalues are 
shown in Fig.~\ref{fig7} where we compare the spectrum of the 
highest precision of 756 binary digits with lower precision spectra 
of 52, 256 and 512 binary digits. As in Fig.~\ref{fig6} for the triangular 
CNPR, for CNPR we also observe that 
the radius and angular extension of the cloud or circle of 
incorrect Jordan block eigenvalues decreases with increasing 
precision. Despite the lower number of $n_A=2000$ as compared to 
$n_A=8000$ of Fig.~\ref{fig3} the number of accurate eigenvalues with 
756 bit precision is certainly considerably higher. 

\begin{figure}[H]
\begin{center}
\includegraphics[width=0.48\textwidth]{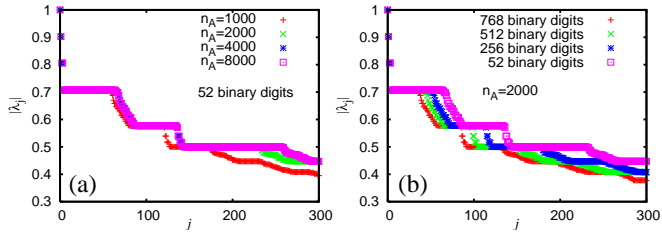}
\end{center}
\caption{(Color online) Modulus $|\lambda_j|$ of the core space 
eigenvalues of $S$ of CNPR, obtained by the Arnoldi 
method, shown versus level number $j$. {\em Panel (a):} data for standard 
double-precision  with 52 binary digits 
with different Arnoldi dimensions $1000\le n_A\le 8000$. 
{\em Panel (b):} data for Arnoldi dimension $n_A=2000$ with 
different numerical precisions between 52 and 768 binary digits. 
\label{fig8}}
\end{figure}

The higher precision Arnoldi method certainly improves  the 
quality of the smaller eigenvalues, e.g. for $|\lambda|<0.3-0.4$, but 
it also implies a strange shortcoming as far as the degeneracies of 
certain particular eigenvalues are concerned. This can be seen in 
Fig.~\ref{fig8} which shows the core space eigenvalues 
$|\lambda_j|$ versus the level number $j$ for various values of the 
Arnoldi dimension and the precision. In these curves we observe flat plateaux 
at certain values $|\lambda_j|=1/\sqrt{n}$ with 
$n=2,\,3,\,4,\,5,\,\ldots$ corresponding to degenerate eigenvalues 
which turn out to be real but with positive or negative values:
$\lambda_j=\pm 1/\sqrt{n}$. 
For fixed standard double-precision arithmetic with 52 binary digits 
the degeneracies increase with increasing Arnoldi dimension and 
seem to saturate for $n_A\ge 4000$. 
However at the given value of $n_A=2000$ the degeneracies {\em decrease} 
with increasing precision of the Arnoldi method. Apparently the higher 
precision Arnoldi method is less able to determine the correct degeneracy 
of a degenerate eigenvalue.

%%%%%%%%%%%%%%%%%%%%%%%%%%%%%%%%%%%%%%%%%%%%%%%%%%%%%%%%%
%%%%%%%%%%%%%%%%%%%%%%%%%%%%%%%%%%%%%%%%%%%%%%%%%%%%%%%%%

This point can be understood as follows. In theory, assuming perfect 
precision, the simple version of Arnoldi method used here (in contrast 
to more complicated block Arnoldi methods) can only determine 
one eigenvector for a degenerate eigenvalue. The reason is that 
for a degenerate eigenvalue we have a particular 
linear combination of the eigenvectors for this eigenvalue which contribute in 
any initial vector (in other words ``one particular'' eigenvector for this 
eigenvalue) and during the Arnoldi iteration this particular eigenvector 
will be perfectly conserved and the generated Krylov space will only contain 
this and no other eigenvector for this eigenvalue.
However, due to round-off errors we obtain at each step new random 
contributions from other eigenvectors of the same eigenvalue and it is only 
due to these round-off errors that we can see the flat plateaux in 
Fig.~\ref{fig8}. Obviously, increasing the precision reduces this round-off 
error effect and the flat plateaux are indeed considerably 
smaller for higher precisions.

The question arises about the origin of the degenerate eigenvalues in the 
core space spectrum. In other examples, such as the WWW for certain 
university networks \cite{univuk}, the degeneracies, especially of the 
leading eigenvalue 1, could be treated by separating and diagonalizing 
the exact subspaces and 
the remaining core space spectrum contained much less or nearly no 
degenerate eigenvalues.
However, here for the CNPR we have ``only'' 
27 subspaces with maximal dimension of 6 containing 71 nodes in total. 
The eigenvalues due to these subspaces are $1,\,-1,\,-0.5,\,0$ with 
degeneracies $27,\,18,\,4,\,22$ (see blue dots in the upper panels 
of Fig.~\ref{fig3}). These exact subspaces exist only due to 
the modest number of future citation links. Even when we take care that 
in all cases the Arnoldi method is applied to the core space without these 
71 subspace nodes, there still remain  a lot of degenerate eigenvalues 
in the core space spectrum.

In Appendix C we explain how the degenerated core space eigenvalues of 
$S$ can be obtained as degenerate subspace eigenvalues of $S_0$ (i.e. 
neglecting the dangling node contributions when determining the invariant 
subspaces). To be precise it turns out that the core space eigenvalues 
of $S$ are decomposed in two groups, the first group is related to 
degenerate subspace eigenvalues of $S_0$ and which can be determined by a 
scheme described in Appendix C, and the second group of eigenvalues 
is given as zeros of a certain rational function (\ref{eq_rationalfunction})
which can be evaluated by the series (\ref{eq_rationseries}) 
which converges only for $|\lambda|>\rho_1$ with $\rho_1\approx 0.902$. 
To determine the zeros of the rational function, outside the range of 
convergence, one can employ an argument of analytical continuation 
using a new method, called ``rational interpolation method'' 
described in detail in Appendix D. Without going into much details here, we 
mention that the main idea of this method is to evaluate this rational 
function at many support 
points on the complex unit circle where the series (\ref{eq_rationseries}) 
converges well and then to use these values to interpolate 
the rational function (\ref{eq_rationalfunction}) by a simpler rational 
function for which the zeros can be determined numerically well even if 
they are inside the unit cercle (where the initial series does not converge). 
For this scheme it is also very important to use high precision computations. 
Typically for a given precision of $p$ binary digits one may chose a certain 
number $n_R$ of eigenvalues to be determined choosing the appropriate 
number of support points (either $2n_R+1$ or $2n_R+2$ depending on the 
variant of the method, see also Appendix D). 
Provided that $n_R$ is not neither too small nor too 
large (depending on the value of $p$) one obtains very reliable core space 
eigenvalues of $S$ of the second group.

\begin{figure}[H]
\begin{center}
\includegraphics[width=0.4\textwidth]{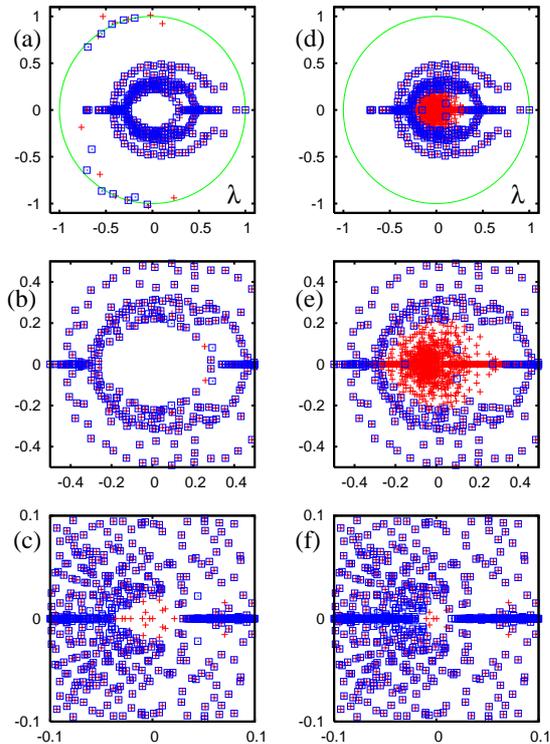}
\end{center}
\caption{(Color online)
{\em  Panel (a):} Comparison of $n_R=340$ core space 
eigenvalues of $S$ for CNPR  obtained by  
two variants of the rational interpolation method (see text) with 
the numerical precision of $p=1024$ binary digits, 
$681$ support points (first variant, red/grey crosses) or 
$682$ support points (second variant, blue/black squares). 
{\em Panel (d):} Comparison of the core space eigenvalues
of CNPR obtained by the high precision Arnoldi method with 
$n_A=2000$ and $p=768$ binary digits 
(red/grey crosses, same data as blue/black squares in Fig. \ref{fig7}) 
with the eigenvalues obtained 
by (both variants of) the rational interpolation method with 
the numerical precision of $p=1024$ binary digits and 
$n_R=300$ eigenvalues (blue/black squares). Here both variants 
with $601$ or $602$ support points provide identical spectra (differences 
below $10^{-14}$). 
{\em Panels (b),(e):} Same as  panels $(a), (d)$ with a zoomed range: 
$-0.5\le \mbox{Re}(\lambda), \mbox{Im}(\lambda)\le 0.5$. 
{\em Panel (c):}  Comparison of the core space spectra 
obtained by the high precision Arnoldi method 
(red/grey crosses, $n_A=2000$ and $p=768$) and by the rational interpolation 
method with $p=12288$, $n_R=2000$ eigenvalues 
(blue/black squares). {\em Panel (f):} Same as $(c)$ 
with $p=16384$, $n_R=2500$ 
for the rational interpolation 
method. Both panels $(c), (f)$ are shown 
in a zoomed range: $-0.1\le \mbox{Re}(\lambda), \mbox{Im}(\lambda)\le 0.1$. 
Eigenvalues outside the shown range coincide up to graphical precision 
and both variants of the rational interpolation method provide 
numerically identical spectra. 
\label{fig9}}
\end{figure}

For example, as can be seen in Fig.~\ref{fig9}, 
for $p=1024$ we obtain $n_R=300$ eigenvalues for which 
the big majority coincides numerically (error $\sim 10^{-14}$) 
with the eigenvalues obtained from the high precision Arnoldi method 
for $768$ binary digits and furthermore both variants of the 
rational interpolation method provide identical spectra. 

However for $n_R=340$ some of the zeros do not coincide with eigenvalues of 
$S$ and most of these deviating zeros lie close to the unit circle. We 
can even somehow distinguish between ``good'' 
zeros (associated to eigenvalues of $S$) being identical for both 
variants of the method and ``bad'' artificial zeros which are completely 
different for both variants (see Fig.~\ref{fig9}). We note that for the 
case of too large $n_R$ values the artificial zeros are extremely sensitive 
to numerical round-off errors (in the high precision variables) and that 
they change strongly, when slightly modifying the support points 
(e.g. a random modification $\sim 10^{-18}$ or simply changing their order in the interpolation 
scheme) or when changing the precise numerical algorithm (e.g. 
between direct sum or Horner scheme for the evaluation of the series of the 
rational function). Furthermore, they do not respect the symmetry that 
the zeros should come in pairs of complex conjugate numbers in case 
of complex zeros. This is because Thiele's rational interpolation 
scheme breaks the symmetry due to complex conjugation once round-off 
errors become relevant.

However, we have carefully verified that for the proper values of $n_R$ not 
being too large (e.g. $n_R=300$ for $p=1024$) the obtained zeros are 
numerically identical (with 52 binary digits in the final result) with respect 
to small changes of the support points (or their order) or with respect 
to different numerical algorithms and that they respect perfectly 
the symmetry due to complex conjugation. 

This method, despite the necessity of high precision calculations, 
is not very expensive, especially for the memory usage, if compared 
with the high precision Arnoldi method. 
Furthermore, its efficiency for the computation time 
can be improved by the trick of 
summing up the largest terms in the series (\ref{eq_rationseries}) 
as a geometrical series which allows to reduce the cutoff value of $l$ 
by a good factor $3$, i.e. replacing $\rho_1\approx 0.902$ by 
$\rho_2=1/\sqrt{2}\approx 0.707$ in the estimate (\ref{eq_lcutoff}) of $l$ 
which gives $l\approx 2\,p+{\rm const.}$ 
We have increased the number of binary digits up to $p=16384$ and we find that 
for $p=1024,2048,4096,6144,8192,12288,16384$ we may use 
$n_R=300,500,900,1200,1500,2000,2500$ 
and still avoid the appearance of artificial zeros. In Fig.~\ref{fig9} 
we also compare the result of the highest precisions $p=12288$ (and $p=16384$)
using $n_R=2000$ ($n_R=2500$) 
with the high precision Arnoldi method with $n_A=2000$ and $p=768$ and these 
spectra coincide well apart from a minor number of smallest 
eigenvalues. 
In general, the complex isolated eigenvalues converge very well 
(with increasing values of $p$ and $n_R$) while the strongly clustered 
eigenvalues on the real axis have more difficulties to converge. 
Comparing the results between $n_R=2000$ and $n_R=2500$ we see that 
the complex eigenvalues coincide on graphical precision 
for $|\lambda|\ge 0.04$ and the 
real eigenvalues for $|\lambda|\ge 0.1$. The Arnoldi method has even more 
difficulties on the real axis (convergence roughly for $|\lambda|\ge 0.15$)
since it has implicitly  to take care of the highly degenerate 
eigenvalues of the first group and for which it has difficulties to 
correctly find the degeneracies (see also Fig.~\ref{fig8}). 

\begin{figure}[H]
\begin{center}
\includegraphics[width=0.46\textwidth]{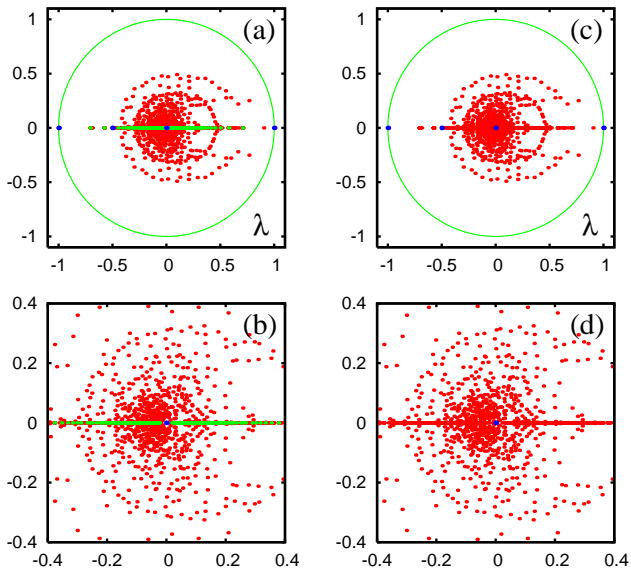}
\end{center}
\caption{ (Color online)
The most accurate spectrum of eigenvalues 
of $S$ for CNPR. {\em Panel (a):} 
red/grey dots represent the core space 
eigenvalues obtained by the rational interpolation method with 
the numerical precision of $p=16384$ binary digits, 
$n_R=2500$ eigenvalues; green (light grey)  dots on $y=0$ axis show 
the degenerate subspace eigenvalues of the matrix $S_0$ 
which are also 
eigenvalues of $S$ with a degeneracy reduced by one (eigenvalues of the 
first group, see text); blue/black dots show the direct 
subspace eigenvalues of $S$ (same as blue/black dots in left upper panel 
in Fig.~\ref{fig3}). 
{\em Panel (c):} red/grey dots represent the core space 
eigenvalues obtained by the high 
precision Arnoldi method with  $n_A=2000$ and 
the numerical precision of $p=768$ 
binary digits and blue dots show the direct 
subspace eigenvalues of $S$. Note that 
the Arnoldi method determines implicitly also the degenerate subspace 
eigenvalues of $S_0$ which are therefore not shown in another color. 
{\em Panels (b), (d)}:  same as in top panels $(a), (c)$
 with a zoomed range: 
$-0.4\le \mbox{Re}(\lambda), \mbox{Im}(\lambda)\le 0.4$. 
\label{fig10}}
\end{figure}

In Fig.~\ref{fig10} we show as summary 
the highest precision spectra of $S$ with core space eigenvalues 
obtained by the Arnoldi method or the rational interpolation method 
(both at best parameter choices) and 
also taking into account the direct subspace eigenvalues of $S$ and the 
above determined eigenvalues of the first group (degenerate subspace 
eigenvalues of $S_0$).

\section{Fractal Weyl law for CNPR} 

The concept of the fractal Weyl law \cite{weyl1,weyl2},\cite{weyl3}
states that the number of states 
$N_\lambda$ in a ring of complex eigenvalues with
$\lambda_c \leq |\lambda| \leq 1$ scales
in a polynomial way with the growth of matrix size:
\begin{equation}
 N_\lambda= a N^b \; .
\label{eqweyl} 
\end{equation} 
where the exponent $b$ is related to the fractal dimension of
underlying invariant set $d_f=2b$. The fractal Weyl law was first discussed
for the problems of quantum chaotic scattering 
in the semiclassical limit \cite{weyl1,weyl2},\cite{weyl3}.
Later it was shown that this law also works for the Ulam 
matrix approximant of the Perron-Frobenius operators
of dissipative chaotic systems with strange attractors
\cite{zhirovulam,ermannweyl}. In \cite{linux} it
was established that the time growing Linux Kernel network 
is also characterized by the fractal Weyl law with the fractal
dimension $d_f \approx 1.3$. 

\begin{figure}[H]
\begin{center}
\includegraphics[width=0.48\textwidth]{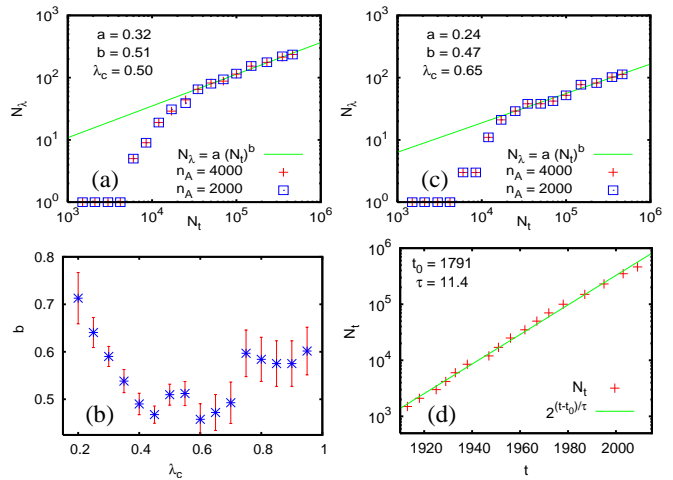}
\end{center}
\caption{(Color online)
Data for 
the whole CNPR at different moments of time.
{\em Panel (a) (or (c)):} 
shows the number $N_\lambda$ of eigenvalues 
with $\lambda_c\leq \lambda\leq 1$
for $\lambda_c=0.50$ (or $\lambda_c=0.65$) versus the effective 
network size $N_t$ where the nodes with publication times after a 
cut time $t$ are removed from the network. 
The green/grey line shows the fractal Weyl law $N_\lambda=a\,(N_t)^b$ 
with parameters $a=0.32\pm 0.08$ ($a=0.24\pm 0.11$)
and $b=0.51\pm 0.02$ ($b=0.47\pm 0.04$) obtained from a fit 
in the range $3\times 10^4\le N_t < 5 \times 10^5$. 
The number $N_\lambda$ includes both 
exactly determined invariant subspace eigenvalues 
and core space eigenvalues obtained from the 
Arnoldi method with double-precision (52 binary 
digits) for $n_A=4000$ (red/grey crosses) and $n_A=2000$ (blue/black squares). 
{\em Panel (b):}  exponent $b$ with error bars 
obtained from the fit $N_\lambda=a\,(N_t)^b$ in the 
range $3\times 10^4\le N_t < 5 \times 10^5$ versus cut value $\lambda_c$. 
{\em Panel (d):} effective network size $N_t$ versus cut time $t$ (in years). 
The green/grey line shows the exponential fit $2^{(t-t_0)/\tau}$ with 
$t_0=1791\pm 3$ and $\tau=11.4\pm 0.2$ representing the number of years 
after which the size of the network (number of papers published 
in all Physical Review journals) is effectively doubled. 
\label{fig11}}
\end{figure}

The fact that $b<1$ implies that
the majority of eigenvalues drop to zero.
We see that this property also appears for the CNPR 
if we test here the validity of the fractal Weyl law by considering 
a time reduced CNPR of size $N_t$ including the $N_t$ papers published 
until the time $t$ (measured in years) for different times $t$ 
in order to obtain a scaling behavior of $N_\lambda$ as a function of $N_t$. 
The data presented in Fig.~\ref{fig11}
shows that the network size grows 
approximately exponentially as $N_t = 2^{(t-t_0)/\tau}$
with the fit parameters $t_0=1791$, $\tau=11.4$. 
The time interval considered in Fig.~\ref{fig11} is 
$1913 \leq t \leq 2009$ since the first data point corresponds 
to $t=1913$ with $N_t=1500$ papers published between 1893 and 1913. 
The results for $N_\lambda$ show that its growth
is well described by the relation $N_\lambda=a\,(N_t)^b$ 
for the range when the number of articles becomes
sufficiently large $3\times 10^4\le N_t < 5 \times 10^5$.
This range is not very large and probably due to that
there is a certain dependence of the exponent $b$ on the range 
parameter $\lambda_c$. At the same time
we note that the maximal matrix size $N$
studied here is probably the largest one used in numerical
studies of the fractal Weyl law.
We have $0.47 < b <0.6$ 
for all $\lambda_c\ge 0.4$ that is definitely smaller than unity
and thus the fractal Weyl law
is well applicable to the CNPR. The value of $b$ increases up to $0.7$ 
for the data points with $\lambda_c<0.4$ but this is due to 
the fact here $N_\lambda$ also includes some numerically incorrect 
eigenvalues related to the numerical instability of the Arnoldi method 
at standard double-precision (52 binary digits) as discussed 
in the beginning of the previous section.

We think that the most appropriate choice for the 
description of the data is obtained at $\lambda_c=0.4$
which from one side excludes small, partly numerically incorrect, values of 
$\lambda$ and on the other side gives sufficiently large
values of $N_\lambda$. Here we have $b=0.49 \pm 02$ corresponding to the
fractal dimension $d=0.98\pm 0.04$. Furthermore, for 
$0.4\le \lambda_c\le 0.7$ we have a rather constant value $b\approx 0.5$ 
with $d_f\approx 1.0$. Of course, it would be
interesting to extend this analysis to a larger
size $N$ of CNPR but for that we still should wait
about 10 years until the network size will be doubled
comparing to the size studied here.

\section{Properties of eigenvectors}

The results for the eigenvalue spectra of CNPR presented in   
the previous sections show that most of the visible eigenvalues 
on the real axis (except for the largest one) 
in Figs.~\ref{fig9} and \ref{fig10} 
are due to the effect of future citations. They appear either directly 
due to $2\times 2$ subblocks 
of the type (\ref{eq_2x2blocks}) with a cycle where 
two papers mutually cite each other giving the degenerate eigenvalues 
of the first group, or indirectly by eigenvalues of the second group which 
are also numerous  on the real axis. On the other hand, as 
can be seen in Fig.~\ref{fig6}, for the 
triangular CNPR, where all future citations 
are removed, there is only the leading eigenvalue $\lambda=1$ and 
a small number of negative eigenvalues with $-0.27<\lambda<0$ on the 
real axis. All other eigenvalues are complex and a considerable number 
of the largest ones are 
relatively close to corresponding complex eigenvalues for 
the whole CNPR with future citations. 

The appearance of future citations is quite specific and is not a typical 
situation for citation networks. 
Therefore we consider the eigenvectors of complex eigenvalues 
for the triangular CNPR which indeed represent 
the typical physical situation without future citations. 
There is no problem to evaluate these eigenvectors by the Arnoldi method, 
either with double-precision, provided the eigenvalue of the eigenvector 
is situated in the region of numerically accurate eigenvalues, or with 
the high precision variant of the Arnoldi method. However, for the triangular 
CNPR we have, according to the semi-analytical theory presented 
above, the explicit formula:
\begin{equation}
\label{eigvec_triangular}
\psi\propto (\lambda\openone-S_0)^{-1}\,e/N=
\sum_{j=0}^{l-1} \lambda^{-(1+j)}\,S_0^j e/N
\end{equation}
where the normalization is given by $\sum_i |\psi(i)|=1$. This 
expression is quite convenient and we  verified that it provides the 
same eigenvectors (up to numerical errors) as the Arnoldi method.

\begin{figure}[H]
\begin{center}
\includegraphics[width=0.48\textwidth]{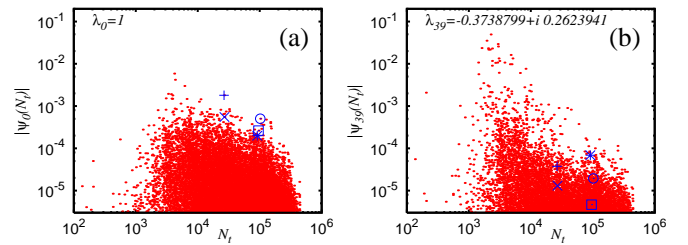}
\end{center}
\caption{(Color online)
Two eigenvectors of the matrix $S$ for the triangular 
CNPR. Both panels show the modulus 
of the eigenvector components $|\psi_j(N_t)|$ versus the time index 
$N_t$ (as used in Fig.~\ref{fig11}) 
with nodes/articles ordered by the publication time (small red/grey dots). 
The blue/black points represent five particular articles: 
BCS 1957 ($+$), Anderson 1958 ($\times$), Benettin et al. 1976 ($*$), 
Thouless 1977 ($\boxdot$) and Abrahams et al. 1979 ($\odot$).
%BCS 1957, Anderson 1958, Benettin et al. 1976, 
%Thouless 1977 and Abrahams et al. 1979.
The left (right) panel corresponds to the real (complex) eigenvalue 
$\lambda_0=1$ ($\lambda_{39}=-0.3738799+i\, 0.2623941$). 
\label{fig12}}
\end{figure}

In Fig.~\ref{fig12} we show two eigenvectors of $S$: 
one $\psi_0$ for the leading eigenvalue 
$\lambda_0=1$ and another $\psi_{39}$
for a complex eigenvalue at $|\lambda_{39}|<1$. 
The eigenvector of $\lambda_0$ gives the PageRank probability
for the triangular CNPR (at $\alpha=1$). We also consider 
the eigenvector for the complex 
eigenvalue $\lambda_{39}=-0.3738799+i\, 0.2623941$
(eigenvalues are ordered by their absolute values
starting from $\lambda_0=1$). In this figure 
the modulus of $|\psi_j(N_t)|$ is shown versus the time index $N_t$ 
as introduced in Fig.~\ref{fig11}. 
We also indicate the positions of five famous  papers: 
BCS 1957 \cite{bcs} at $K=6$, 
Anderson 1958 \cite{anderson} $K=63$, 
Benettin et al. 1976 \cite{benettin} $K=441$, 
Thouless 1977 \cite{thouless} $K=256$ and 
Abrahams et al. 1979 \cite{abrahams} $K=74$. 
In the first eigenvector 
for $\lambda_0=1$ all of these papers have quite dominating positions, 
especially BCS 1957 and Abrahams et al. 1979 
which are the most important ones 
if compared to papers of comparable publication date. Only considerably 
older papers have higher positions in this vector.

For the second eigenvector with complex eigenvalue the older papers 
(with $10^3<N_t<10^4$ corresponding to publications times between 
1910 and 1940) are strongly enhanced in its importance while 
the above five  
famous papers lose their importance. 
The top 3 positions of largest amplitude $|\psi_{39}(i)|$
correspond to DOI 10.1103/PhysRev.14.409 (1919),
10.1103/PhysRev.8.561 (1916), 10.1103/PhysRev.24.97 (1917).
These old articles study 
the radiating potentials of nitrogen, ionization impact in gases
and  the abnormal low voltage arc. It is clear that 
this eigenvector selects
a certain community of old articles
related to a certain ancient field of interest.
This fact is in agreement with the studies of eigenvectors
of Wikipedia network \cite{wikileo} showing that the eigenvectors
with $0 < |\lambda| <1$ select specific communities.

It is interesting to note that the top node of the vector $\psi_0$
appears in the position $K_{39}=39$ in local rank index 
of the vector $\psi_{39}$ (ranking in decreasing order by modulus of 
$|\psi(i)|$).
On the other side the top node of $\psi_{39}$ appears at position
$K_0=30$ of vector $\psi_0$. This illustrates how different nodes 
contribute to different eigenvectors of $S$.

It is useful to characterize the eigenvectors by their 
Inverse Participation Ratio (IPR)
$\xi_i=(\sum_j |\psi_i(j)|^2)^2/\sum_j |\psi_i(j)|^4$
which gives an effective number of nodes
populated by an eigenvector $\psi_i$ (see e.g. \cite{ggs1,wikileo}).
For the above two vectors we find
$\xi_0=  20.67$ and $\xi_{39}= 10.76$. This means
that $\xi_{39}$ is mainly located on approximately $11$ nodes.
For $\xi_0$ this number is twice larger in agreement with
data of Fig.~\ref{fig12} which show a clearly broader
distribution comparing to $\xi_{39}$.

We also considered a few tens of eigenstates of $S$ of the whole CNPR.
They are mainly located on the complex plane around the largest oval
curve well visible in the spectrum (see Fig.~\ref{fig10} top right panel).
The IPR value of these eigenstates with $|\lambda| \sim 0.4$
varies in the range $4 < \xi <13$ showing that they are located on some 
effective quasi-isolated
communities of articles. About $10$ of them are
related to the top article of $\psi_{39}$
shown in Fig.~\ref{fig12} meaning that these ten vectors 
represent various linear combinations of vectors
 on practically the same community. In global, we can say
that the eigenstates of $G$ are well localized since $\xi \ll N$.
A similar situation was seen for the Wikipedia network
\cite{wikileo}.

Of course, in addition to $\xi$ it is also useful  to 
consider the whole distribution of $\psi$ 
amplitudes over the nodes. Such a consideration has been done for
the Wikipedia network in \cite{wikileo}.
For the CNPR we leave such detailed studies for further
investigations.

\section{CheiRank versus PageRank for CNPR}

\begin{figure}[H]
\begin{center}
\includegraphics[width=6.2cm,angle=-90]{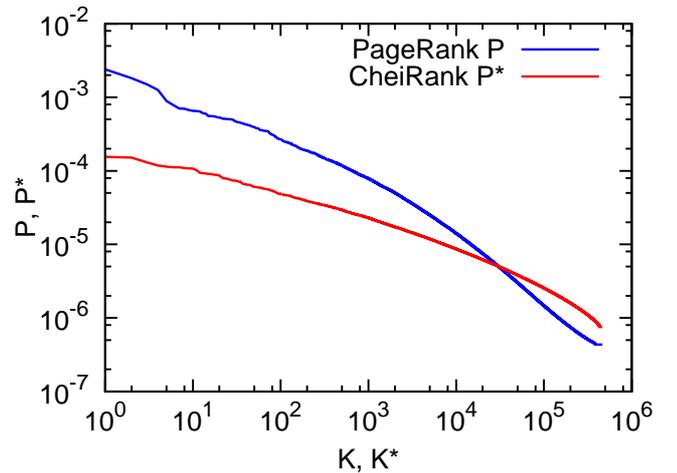}
\end{center}
\caption{(Color online) Dependence
of probability of Page\-Rank $P$ (CheiRank $P^*$) 
on corresponding index $K$ ($K^*$) for the 
CNPR at $\alpha=0.85$.
\label{fig13}}
\end{figure}

The dependence of PageRank probability $P(K)$
on PageRank index $K$ is shown in Fig.~\ref{fig13}.
The results are similar to those of 
\cite{redner,maslov}, \cite{fortunato,fortunato2}.
We note that the PageRank of the triangular CNPR has the same top 9 
articles as for the whole CNPR (both at $\alpha=0.85$ and with 
a slight interchanged  order of positions $7$, $8$, $9$).
This confirms that the future citations produce only a small effect
on the global ranking. 

Following previous studies \cite{alik},\cite{zzswiki,2dmotor},
in addition to the Google matrix $G$ we also construct the matrix
$G^*$ following the same definition (\ref{eq1})
but for the network with inverted direction of links.
The PageRank vector of this matrix $G^*$ is called the 
CheiRank vector with probability  $P^*(K^*_i)$ 
and CheiRank index $K^*$. The dependence of  $P^*(K^*_i)$
is shown in Fig.~\ref{fig13}. We find that the IPR values of 
$P$ and $P^*$ are $\xi=59.54$ and $1466.7$ respectively.
Thus $P^*$ is extended over significantly larger 
number of nodes comparing to $P$. 
A power law fit of the decay 
$P \propto 1/K^\beta$, $P^* \propto 1/{K^*}^\beta$,
done for a range $K, K^* \leq 2 \times 10^5$
gives $\beta \approx 0.57$ for $P$ and $\beta \approx 0.4$ for $P^*$.
However, this is only an approximate description since there is
a visible curvature (in a double logarithmic representation) in these 
distributions. The corresponding 
frequency distributions of ingoing links
have exponents $\mu = 2.87$ while the 
distribution of outgoing links has $\mu \approx 3.7$
for outdegree $k \ge 20$,
even if the whole frequency dependence
in this case is rather curved 
and a power law fit is rather
approximate in this case.
Thus the usual relation $\beta = 1/(\mu -1 )$ \cite{googlebook,ggs1,zzswiki}
approximately works.
\begin{figure}[H]
\begin{center}
\includegraphics[width=0.48\textwidth]{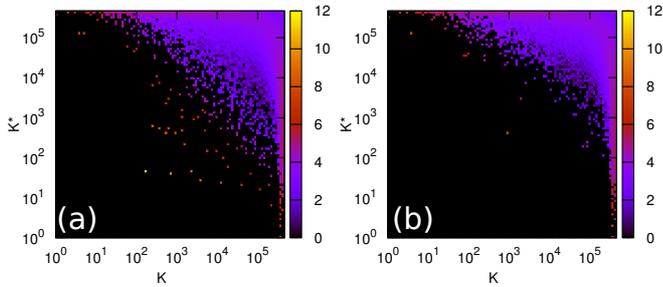}
\end{center}
\caption{(Color online) Density 
distribution $W(K,K^{*}) = dN_{i}/dKdK^{*}$
of Physical Review articles in the
PageRank-CheiRank plane $(K,K^*)$.
Color bars show the natural logarithm of density, changing
from minimal nonzero density (dark) to maximal one (white),
zero density is shown by black.
Panel $(a)$: all articles of CNPR;
panel $(b)$: CNPR without Rev. Mod. Phys.
\label{fig14}}
\end{figure}

The correlation between PageRank and CheiRank vectors
can be characterized by the correlator
$\kappa=N \sum_{i=1}^{N} P(i) P^*(i) - 1$ \cite{alik,2dmotor}.
Here we find $\kappa=-0.2789$ for all CNPR,
and $\kappa=-0.3187$ for CNPR without Rev. Mod. Phys.
This is the most strong negative value of $\kappa$
among all directed networks studied previously \cite{2dmotor}.
In a certain sense the situation is somewhat similar to
the Linux Kernel network where $\kappa \approx 0$ or slightly negative
($\kappa > -0.1$ \cite{alik}). 
For CNPR, we can say that due to a almost triangular structure 
of $G$ and $G^*$ there is a very little overlap of
top ranking in $K$ and $K^*$ that leads to a negative
correlator value, since the components 
$ P(i) P^*(i)$ of the sum for $\kappa$ are small.

Each article $i$ has two indexes $K_i,K^*_i$
so that it is convenient to see their distribution on
2D PageRank-CheiRank plane. The density distribution 
$W(K,K^{*}) = dN_{i}/dKdK^{*}$ is shown in Fig.~\ref{fig14}.
It is obtained from $100 \times 100$ cells 
equidistant in log-scale
(see details in \cite{zzswiki,2dmotor}). 
For the CNPR the density is homogeneous along lines $K=-K^*+const$
that corresponds to the absence of correlations between
$P$ and $P^*$ \cite{zzswiki,2dmotor}.
For the CNPR without Rev. Mod. Phys. we have an additional 
suppression of density at low $K^*$ values. Indeed,
Rev. Mod. Phys. contains mainly review articles with a large number
of citations that place them on top of CheiRank. 
At the top 3 positions of $K^*$ of CNPR we have
DOI 10.1103/PhysRevA.79.062512, 10.1103/PhysRevA.79.062511,
10.1103/RevModPhys.81.1551 of 2009.
These are articles with long citation lists on
 $K$ shell diagram  4d transition elements;
hypersatellites of 3d transition metals;
superconducting phases of $f$ electron compounds.
For CNPR without Rev. Mod. Phys. the first two articles are the same
and the third one has DOI 10.1103/PhysRevB.80.224501 being about
model for the coexistence of d wave 
superconducting and charge density wave order in
in high temperature cuprate superconductors.
We see that the most recent articles with long citation lists are dominating.

The top PageRank articles are analyzed in detail in
\cite{redner,maslov},\cite{fortunato,fortunato2} 
and we do not discuss them here.

It is also useful to consider two-dimensional rank 2DRank $K_2$
defined by counting nodes in order of their appearance on
ribs of squares in $(K,K^*)$ plane with the square size growing from
$K=1$ to $K=N$ \cite{zzswiki}. It selects highly cited articles with 
a relatively long citation list. For CNPR, we have top 3 such articles
with DOI 10.1103/RevModPhys.54.437 (1982),
10.1103/RevModPhys.65.851 (1993), 10.1103/RevModPhys.58.801 (1986).
Their topics are electronic properties of two dimensional systems,
pattern formation outside of equilibrium,  spin glasses  facts and concepts.
The 1st one located at $K=183$, $K^*=49$ is well visible
in the left panel of Fig.~\ref{fig14}. For CNPR without Rev. Mod. Phys.
we find at $K_2=1$ the article with DOI 10.1103/PhysRevD.54.1 (1996)
entitled  {\it Review of Particle Physics} with a lot of information
on physical constants.

For the ranking of articles about persons in Wikipedia networks
\cite{zzswiki,wikievol},\cite{cultwiki}, PageRank, 2DRank, CheiRank highlights
in a different manner various sides of human activity.
For the CNPR, these 3 ranks also select different types of articles,
however, due a  triangular structure of $G, G^*$ and
absence of correlations between PageRank and CheiRank
vectors the useful side of 2DRank and CheiRank remains less evident.

\section{ImpactRank for influence propagation}

It is interesting to quantify how an influence of a given article propagates 
through the whole CNPR. To analyze this property we consider the 
following propagator acting on an initial vector $v_0$ located on a given 
article:
\begin{equation}
%   v_f= [(1-\gamma)/(1-\gamma G)]v_0 \; ,  \; 
%v^*_f= [(1-\gamma)/(1-\gamma G^*)]v_0 \; .
v_f=\frac{1-\gamma}{1-\gamma G}\,v_0\quad,\quad 
v^*_f=\frac{1-\gamma}{1-\gamma G^*}\,v_0\ .
\label{eqimpact} 
\end{equation} 
Here $G, G^*$ are the Google matrices defined above,
$\gamma$ is a new  impact damping factor being in a range
$\gamma \sim 0.5 - 0.9$, $v_f$ in the final vector
generated by the propagator (\ref{eqimpact}).
This vector is normalized to unity $\sum_i v_f(i)=1$ and one can easily 
show that it is equal to the PageRank vector of a modified Google matrix 
given by
\begin{equation}
\label{eq_google_modified}
\tilde G=\gamma\,G+(1-\gamma)\,v_0\,e^T
\end{equation}
where $e$ is the vector with unit elements. This modified Google matrix 
corresponds to a stochastic process where at a certain time 
a given probability distribution is propagated 
with probability $\gamma$ using the initial Google matrix $G$ 
and with probability $(1-\gamma)$ the probability 
distribution is reinitialized with the vector $v_0$. Then 
$v_f$ is the stationary vector from this stochastic process. 
Since the initial Google matrix $G$ has a similar form, 
$G=\alpha S+(1-\alpha)e\,e^T/N$ with the damping factor $\alpha$, 
the modified Google matrix can also be written as:
\begin{equation}
\label{eq_google_modified2}
\tilde G=\tilde\alpha\,S+(1-\tilde\alpha)\,v_p\,e^T\quad,\quad
\tilde\alpha=\gamma\alpha\quad,
\end{equation}
with the personalization vector \cite{googlebook} 
\begin{equation}
\label{eq_personalization_vector}
v_p=\frac{\gamma(1-\alpha)e/N+(1-\gamma)v_0}{1-\gamma\alpha}
\end{equation}
which is also sum normalized: $\sum_i v_p(i)=1$. Obviously similar 
relations hold for $G^*$ and $v_f^*$. 
\begin{figure}[H]
\begin{center}
\includegraphics[width=2.7cm, angle=-90]{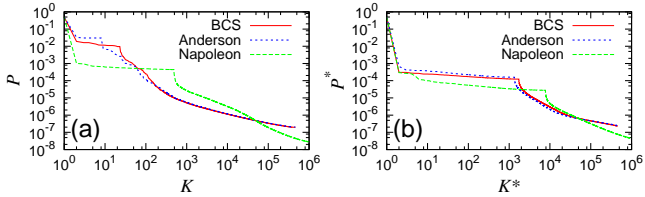}
\end{center}
\caption{(Color online) 
Dependence of impact vector $v_f$  probability
 $P$ and $P^*$
($(a)$ and $(b)$ panels) on the corresponding
ImpactRank index $K$ and $K^*$ for
an initial article $v_0$  as BCS \cite{bcs} and
Anderson \cite{anderson} in CNPR,
and Napoleon in English Wikipedia network 
from \cite{cultwiki}.
Here the impact damping factor is $\gamma=0.5$.
\label{fig15}}
\end{figure}

The relation (\ref{eqimpact})
can be viewed as a Green function with damping $\gamma$.
Since $\gamma<1$ the expansion in a geometric series 
is convergent and $v_f$ can be obtained from about
$200$ terms of the expansion for $\gamma \sim 0.5$. 
The stability of $v_f$ is verified by changing the number of 
terms. The obtained vectors $v_f$, $v^*_f$ can be considered
as effective PageRank, CheiRank probabilities 
$P$, $P^*$ and all nodes can be ordered
in the corresponding rank index $K$, $K^*$,
which we will call ImpactRank.

The results for 2 initial vectors located on 
BCS \cite{bcs} and Anderson \cite{anderson} articles
are shown in Fig.~\ref{fig15}. In addition
we show the same probability for the Wikipedia article
{\it Napoleon} for the English Wikipedia 
network analyzed in \cite{cultwiki}. 
The direct analysis of the distributions shows that
the original article is located at the top position,
the next step like structure corresponds to
the articles reached by first outgoing (ingoing) 
links from $v_0$ for $G$ ($G^*$). The next visible step
correspond to a second link step.

\begin{table*}[!ht]
\caption{Spreading of impact on "Theory of superconductivity"
paper by "J. Bardeen, L. N. Cooper and J. R. Schrieffer
(doi:10.1103/PhysRev.108.1175) by Google matrix $G$ with
$\alpha=0.85$
 and $\gamma=0.5$ }
\begin{center}
\resizebox{15 cm}{!}{
\begin{tabular}{|c|c|c|}
  \hline
  ImpactRank & DOI & Title of paper \\
  \hline
1 & 10.1103/PhysRev.108.1175 & Theory of superconductivity \\
2 & 10.1103/PhysRev.78.477 & Isotope effect in the superconductivity of mercury \\
3 & 10.1103/PhysRev.100.1215 & Superconductivity at millimeter wave frequencies\\
4 & 10.1103/PhysRev.78.487 & Superconductivity of isotopes of mercury \\
5 & 10.1103/PhysRev.79.845 & Theory of the superconducting state.
i. the ground \ldots \\ %state at the absolute zero of temperature \\
6 & 10.1103/PhysRev.80.567 & Wave functions for superconducting electrons \\
7 & 10.1103/PhysRev.79.167 & The hyperfine structure of ni$^{61}$ \\
8 & 10.1103/PhysRev.97.1724 & Theory of the Meissner effect in superconductors \\
9 & 10.1103/PhysRev.81.829 & Relation between lattice vibration
and London \ldots \\ %theories of superconductivity \\
10 & 10.1103/PhysRev.104.844 & Transmission of superconducting
films \ldots \\ %at millimeter-microwave and far infrared frequencies \\
  \hline
\end{tabular}}
\end{center}
\end{table*}

\begin{table*}[!ht]
\caption{Spreading of impact on "Absence of diffusion in certain
random lattices" paper by P. W. Anderson
(doi:10.1103/PhysRev.109.1492) by Google matrix $G$. with
$\alpha=0.85$ and $\gamma=0.5$ }
\begin{center}
\resizebox{15 cm}{!}{
\begin{tabular}{|c|c|c|}
  \hline
  ImpactRank & DOI & Title of paper \\
  \hline
1 & 10.1103/PhysRev.109.1492 & Absence of diffusion in certain random lattices \\
2 & 10.1103/PhysRev.91.1071 & Electronic structure of f centers: saturation of \ldots \\ %the electron spin resonance\\
3 & 10.1103/RevModPhys.15.1 & Stochastic problems in physics and astronomy \\
4 & 10.1103/PhysRev.108.590 & Quantum theory of electrical transport phenomena \\
5 & 10.1103/PhysRev.48.755 & Theory of pressure effects of foreign gases on spectral lines \\
6 & 10.1103/PhysRev.105.1388 & Multiple scattering by quantum-mechanical systems \\
7 & 10.1103/PhysRev.104.584 & Spectral diffusion in magnetic resonance \\
8 & 10.1103/PhysRev.74.206 & A note on perturbation theory \\
9 & 10.1103/PhysRev.70.460 & Nuclear induction \\
10 & 10.1103/PhysRev.90.238 & Dipolar broadening of magnetic resonance lines \ldots \\
%in magnetically diluted crystals \\
  \hline
\end{tabular}}
\end{center}
\end{table*}

\begin{table*}[!ht]
\caption{Spreading of impact on "Theory of superconductivity"
paper by "J. Bardeen, L. N. Cooper and J. R. Schrieffer
(doi:10.1103/PhysRev.108.1175) by Google matrix $G^*$ with
$\alpha=0.85$
 and $\gamma=0.5$ }
\begin{center}
\resizebox{16 cm}{!}{
\begin{tabular}{|c|c|c|}
  \hline
  ImpactRank & DOI & Title of paper \\
  \hline
1 & 10.1103/PhysRev.108.1175 & Theory of superconductivity \\
2 & 10.1103/PhysRevB.77.104510 & Temperature-dependent gap edge in strong-coupling \ldots \\ %superconductors determined using the eliashberg-nambu formalism \\
3 & 10.1103/PhysRevC.79.054328 & Exact and approximate ensemble treatments of thermal \ldots \\ % pairing in a multilevel model \\
4 & 10.1103/PhysRevB.8.4175 & Ultrasonic attenuation in superconducting molybdenum \\
5 & 10.1103/RevModPhys.62.1027 & Properties of boson-exchange superconductors \\
6 & 10.1103/PhysRev.188.737 & Transmission of far-infrared radiation through thin films \ldots \\ % of superconducting amorphous bismuth and gallium and beta-phase gallium \\
7 & 10.1103/PhysRev.167.361 & Superconducting thin film in a magnetic field - theory of \ldots \\ % nonlocal  and nonlinear effects. i. specular reflection \\
8 & 10.1103/PhysRevB.77.064503 & Exact mesoscopic correlation functions of the Richardson \ldots \\ %pairing  model \\
9 & 10.1103/PhysRevB.10.1916 & Magnetic field attenuation by thin superconducting lead films \\
10 & 10.1103/PhysRevB.79.180501 & Exactly solvable pairing model for superconductors with \ldots \\ % p$_x$+ip$_y$-wave symmetry \\
  \hline
\end{tabular}}
\end{center}
\end{table*}

\begin{table*}[!ht]
\caption{Spreading of impact on "Absence of diffusion in certain
random lattices" paper by P. W. Anderson
(doi:10.1103/PhysRev.109.1492) by Google matrix $G^*$. with
$\alpha=0.85$ and $\gamma=0.5$ }
\begin{center}
\resizebox{16 cm}{!}{
\begin{tabular}{|c|c|c|}
  \hline
  ImpactRank & DOI & Title of paper \\
  \hline
1 & 10.1103/PhysRev.109.1492 & Absence of diffusion in certain random lattices \\
2 & 10.1103/PhysRevA.80.053606 & Effects of interaction on the diffusion of atomic \ldots \\ % matter waves in one-dimensional quasiperiodic potentials \\
3 & 10.1103/RevModPhys.80.1355 & Anderson transitions \\
4 & 10.1103/PhysRevE.79.041105 & Localization-delocalization transition in hessian \ldots \\ % matrices of topologically disordered systems \\
5 & 10.1103/PhysRevB.79.205120 & Statistics of the two-point transmission at \ldots \\ %  anderson localization transitions \\
6 & 10.1103/PhysRevB.80.174205 & Localization-delocalization transitions \ldots \\ % in a two-dimensional quantum percolation model: von neumann entropy studies \\
7 & 10.1103/PhysRevB.80.024203 & Statistics of renormalized on-site energies and \ldots \\ %  renormalized hoppings for anderson localization in two and three dimensions \\
8 & 10.1103/PhysRevB.79.153104 & Flat-band localization in the Anderson-Falicov-Kimball model \\
9 & 10.1103/PhysRevB.74.104201 & One-dimensional disordered wires with Poschl-Teller potentials \\
10 & 10.1103/PhysRevB.71.235112 & Critical wave-packet dynamics in the power-law bond \ldots \\ % disordered anderson model \\
  \hline
\end{tabular}}
\end{center}
\end{table*}

\begin{table*}[!ht]
\caption{Spreading of impact on the article of "Napoleon" in
English Wikipedia by Google matrix $G$ and $G^*$. with
$\alpha=0.85$ and $\gamma=0.5$ }
\begin{center}
\resizebox{15 cm}{!}{
\begin{tabular}{|c|c|c|}
  \hline
  ImpactRank & Articles ($G$ case) & Articles ($G^*$ case) \\
  \hline
1 & Napoleon & Napoleon \\
2 & French Revolution & List of orders of battle \\
3 & France & Lists of state leaders by year \\
4 & First French Empire & Names inscribed under the Arc de Triomphe \\
5 & Napoleonic Wars & List of battles involving France \\
6 & French First Republic & Order of battle of the Waterloo Campaign \\
7 & Saint Helena & Napoleonic Wars \\
8 & French Consulate & Wagram order of battle \\
9 & French Directory & Departments of France \\
10 & National Convention & Jena-Auerstedt Campaign Order of Battle\\
%in magnetically diluted crystals \\
  \hline
\end{tabular}}
\end{center}
\end{table*}

Top ten articles for these 3 vectors are shown in Tables I, II, III, IV, V.
The analysis of these top articles confirms that they 
are closely linked with the initial
article and thus the ImpactRank gives relatively good ranking results.
At the same time, some questions for such ImpactRanking still remain
to be clarified. For example, in Table IV we find at 
the third position the well known
Rev. Mod. Phys. on Anderson transitions but the paper of Abrahams {\it et al.} 
\cite{abrahams} appears only on far positions $K^* \approx 300$. 
The situation is changed if we consider all CNPR links 
as bi-directional obtaining a non-directional network. 
Then the paper \cite{abrahams} appears on the 
second position directly after initial article \cite{anderson}.
We think that such a problem appears due to triangular structure of CNPR
where there is no intersection of forward and backward flows.
Indeed, for the case of Napoleon we do not see such difficulties.
Thus we hope that such an approach can be applied to other directed networks.

\section{Models of random Perron-Frobenius matrices}
\label{sec_RPFM}

In this section we discuss the spectral properties 
of several random matrix models of Perron-Frobenius operators 
characterized by non-negative matrix elements and column sums normalized 
to unity. We call these models Random Perron-Frobenius Matrices (RPFM). 
To construct these models for a given matrix $G$ of dimension $N$ we 
draw $N^2$ independent matrix elements $G_{ij}\ge 0$ from 
a given distribution $p(G)$ (with $p(G)=0$ for $G<0$) with average 
$\langle G\rangle=1/N$ and finite variance 
$\sigma^2=\langle G^2\rangle-\langle G\rangle^2$. 
A matrix obtained in this way obeys the column sum normalization only 
in average but not exactly for an arbitrary realization. Therefore 
we renormalize all columns to unity after having drawn the matrix elements. 
This renormalization provides some (hopefully small) correlations between 
the different matrix elements. 

Neglecting these correlations for sufficiently 
large $N$ the statistical average of the RPFM is simply given by 
$\langle G_{ij}\rangle=1/N$ which is a projector matrix with the 
eigenvalue $\lambda=1$ of multiplicity 1 
and the corresponding eigenvector being the 
uniform vector $e$ (with $e_i=1$ for all $i$). The other eigenvalue 
$\lambda=0$ is highly degenerate of multiplicity $N-1$ and its eigenspace 
contains all vectors orthogonal to the uniform vector $e$. 
Writing the matrix elements of a RPFM as 
$G_{ij}=\langle G_{ij}\rangle+\delta G_{ij}$ we may consider the 
fluctuating part $\delta G_{ij}$ as a perturbation which only weakly 
modifies the unperturbed eigenvector $e$ for $\lambda=1$ but for 
the eigenvalue $\lambda=0$ we have to apply degenerate perturbation theory 
which requires the diagonalization of $\delta G_{ij}$. 
According to the theory 
of non-symmetric real random Gaussian matrices \cite{mehta,ginibre,sommers} 
it is well established 
that the complex eigenvalue density of such a matrix is uniform on a 
circle of radius $R=\sqrt{N}\sigma$ with $\sigma^2$ being the variance 
of the matrix elements. One can also expect that this holds for more 
general, non-Gaussian, distributions with finite variance provided that 
we exclude extreme long tail distribution where the typical 
values are much smaller than $\sigma$. 
Therefore we expect that the eigenvalue density of a RPFM is determined 
by a single parameter being the variance $\sigma^2$ of the matrix 
elements resulting 
in a uniform density on a circle of radius $R=\sqrt{N}\sigma$ 
around $\lambda=0$, in 
addition to the unit eigenvalue $\lambda=1$ which is always an exact 
eigenvalue due to sum normalization of columns. 

\begin{figure}[H]
\begin{center}
\includegraphics[width=0.48\textwidth]{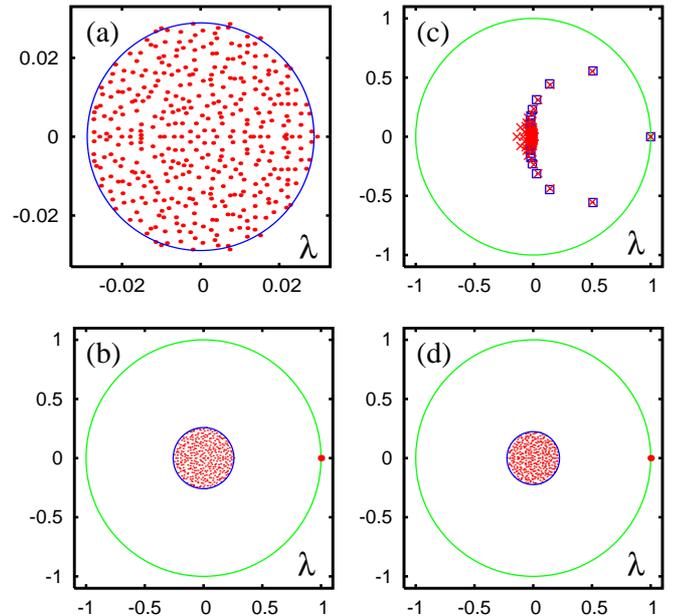}
\end{center}
\vglue -0.3cm
\caption{\label{fig16} (Color online)
Panel $(a)$ shows the  spectrum (red/grey dots) of 
one realization of a full uniform RPFM with 
dimension $N=400$ and matrix elements uniformly distributed in 
the interval $[0,2/N[$; the blue/black circle represents the theoretical 
spectral border with radius $R=1/\sqrt{3N}\approx 0.02887$. 
The unit eigenvalue $\lambda=1$ is not shown due to the zoomed 
presentation range. 
Panel $(c)$ shows the spectrum of 
one realization of triangular RPFM (red/grey crosses) with 
non-vanishing matrix elements 
uniformly distributed in the interval $[0,2/(j-1)[$ 
and a triangular matrix with 
non-vanishing  elements  $1/(j-1)$ (blue/black squares); 
here $j=2,3,\ldots,N$ is the index-number of non-empty columns 
and the first column with $j=1$ corresponds to a dangling node with 
elements $1/N$ for both triangular cases. 
Panels $(b), (d)$ show the complex eigenvalue spectrum (red/grey dots) 
of a sparse RPFM with dimension $N=400$ and $Q=20$ non-vanishing elements 
per column at random positions. Panel $(b)$ (or $(d)$)  corresponds 
to the case of uniformly distributed non-vanishing elements in 
the interval $[0,2/Q[$ (constant non-vanishing elements being $1/Q$);
the blue/black circle represents the theoretical 
spectral border with radius $R=2/\sqrt{3Q}\approx 0.2582$ 
($R=1/\sqrt{Q}\approx 0.2236$). In  panels $(b), (d)$ 
$\lambda=1$ is shown by a larger red dot for better visibility. 
The unit circle is shown by green/grey line (panels $(b), (c), (d)$).
}
\end{figure}

We now consider different variants of RPFM. The first 
variant is a full matrix with each element uniformly distributed 
in the interval $[0,2/N[$ which gives the variance $\sigma^2=1/(3N^2)$
and the spectral radius $R=1/\sqrt{3N}$. The second variant is 
a sparse RPFM matrix with $Q$ non-vanishing elements per column and which are 
uniformly distributed in the interval $[0,2/Q[$. Then the probability 
distribution is given by $p(G)=(1-Q/N)\delta(G)+(Q/N)\,\chi_{[0,2/Q[}(G)$ 
where $\chi_{[0,2/Q[}(G)$ is the characteristic function on the 
interval $[0,2/Q[$ (with values being 1 for $G$ in this interval and 0 
for $G$ outside this interval). The average is indeed $\langle G\rangle=1/N$ 
and the variance is $\sigma^2=4/(3NQ)$ (for $N\gg Q$) providing 
the spectral radius $R=2/\sqrt{3Q}$. 
We may also consider a sparse RPFM where we have exactly 
$Q$ non-vanishing constant elements of value $1/Q$ 
in each column with random positions resulting 
in a variance $\sigma^2=1/(NQ)$ and $R=1/\sqrt{Q}$. The theoretical 
predictions for these three variants of RPFM 
coincide very well with numerical simulations. In Fig.~\ref{fig16}
the complex eigenvalue spectrum for one realization of each of the
three cases is shown for $N=400$ and $Q=20$ clearly confirming the 
circular uniform eigenvalue density with the theoretical values of $R$. 
We  also confirm numerically the scaling behavior of $R$ as a function 
of $N$ or $Q$.

Motivated by the Google matrices of DNA sequences \cite{kandiah}, where 
the matrix elements are distributed with a power law, we also considered 
a power law variant of RPFM with $p(G)=D(1+aG)^{-b}$ for $0\le G\le 1$ 
and with an exponent $2<b<3$. The condition $G\le 1$ is required because 
of the column sum normalization. The parameters 
$D$ and $a$ are determined by normalization and the average 
$\langle G\rangle=1/N$. In the limit $N^{b-2}\gg 1$ we find 
$a\approx N/(b-2)$ and $D\approx N(b-1)/(b-2)$. 
For $b>3$ the variance would scale with $\sim N^{-2}$ resulting in 
$R\sim 1/\sqrt{N}$ as in the first variant with uniformly distributed 
matrix elements. However, for $b<3$ this scaling is different and we find 
(for $N^{b-2}\gg 1$)~:
\begin{equation}
\label{eq_power_radius}
R=C(b)\,N^{1-b/2}\quad,\quad C(b)=(b-2)^{(b-1)/2}\,\sqrt{\frac{b-1}{3-b}}
\quad. 
\end{equation}

\begin{figure}[H]
\begin{center}
\includegraphics[width=0.48\textwidth]{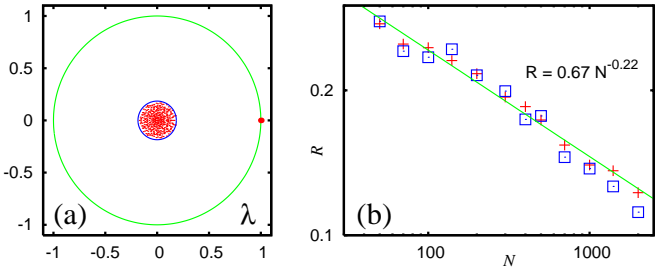}
\end{center}
\vglue -0.3cm
\caption{\label{fig17} (Color online)
Panel $(a)$ shows the  spectrum (red/grey dots) of 
one realization of the power law RPFM with dimension $N=400$ 
and decay exponent $b=2.5$ (see text); the unit eigenvalue 
$\lambda=1$ is shown by a large red/grey dot,
the unit circle is shown by green/grey curve;
the blue/black circle represents the 
spectral border with theoretical radius 
$R= \approx 0.1850$ (see text). 
Panel $(b)$ shows the dependence of the spectrum border radius on 
matrix size $N$ for $50\le N\le 2000$; red/grey crosses represent 
the radius obtained from theory (see text); blue/black squares 
correspond to the spectrum border radius obtained 
numerically from a small number of 
eigenvalues with maximal modulus; the green/grey line shows the 
fit $R=C\,N^{-\eta}$ of red/grey crosses  
with $C=0.67\pm 0.03$ and $\eta=0.22\pm 0.01$. 
}
\end{figure}

Fig.~\ref{fig17} shows the results of numerical diagonalization 
for one realization with $N=400$ and $b=2.5$ such that we expect 
$R\sim N^{-0.25}$. It turns out that the circular eigenvalue density 
is rather well confirmed and the ``theoretical radius'' is indeed given by 
$R=\sqrt{N}\sigma$ if the variance $\sigma^2$ of matrix elements 
is determined by an average over the 
$N^2$ matrix elements of the given matrix. A study for different values 
of $N$ with $50\le N\le 2000$ also confirms the dependence 
$R=C\,N^{-\eta}$ with fit values $C=0.67\pm 0.03$ and $\eta=0.22\pm 0.01$. 
The value of $\eta=0.22$ is close to the theoretical 
value $1-b/2=0.25$ but the prefactor $C=0.67$ is smaller than 
its theoretical value $C(2.5)\approx 1.030$. This is due to the 
correlations introduced by the additional column sum normalization 
after drawing the random matrix elements. 
Furthermore for the power law model with $b<3$ we should not expect a precise 
confirmation of the uniform circular density obtained 
for Gaussian distribution matrix elements. Actually, a more detailed 
numerical analysis 
of the density shows that the density for the power law model is not exactly 
uniform, in particular for values of $b$ close to 2. 

The important observation is that a generic RPFM (full, sparse 
or with power law distributed matrix elements) has a complex eigenvalue 
density rather close to a uniform circle of a quite small radius 
(depending on the parameters $N$, $Q$ or $b$). The fact, that the realistic 
networks (e.g. certain university WWW-networks) have Google matrix spectra 
very different from this \cite{univuk},
shows that in these networks 
there is indeed a subtle network structure and that already slight 
random perturbations or variations immediately result in 
uniform circular eigenvalue spectra. This was already observed 
in \cite{ggs1,ggs2}, where it was shown that certain modest random 
changes in the network links already provide such circular eigenvalue 
spectra. 

We also determine the PageRank for the different 
variants of the RPFM, i.e. the eigenvector for the eigenvalue $\lambda=1$. 
It turns out that it is rather uniform that is rather natural since 
this eigenvector should be close to the uniform vector $e$ which is the 
``PageRank'' for the average matrix $\langle G_{ij}\rangle = 1/N$. This 
also holds when we use a damping factor $\alpha=0.85$ for the RPFM.

Following the above discussion about triangular networks (with 
$G_{ij}=0$ for $i\ge j$) we also study numerically a triangular 
RPFM where for $j\ge 2$ and $i<j$ the matrix elements $G_{ij}$ are 
uniformly distributed in 
the interval $[0,2/(j-1)[$ and for $i\ge j$ we have $G_{ij}=0$. 
Then the first column is empty, that means it 
corresponds to a dangling node and it needs to be replaced by $1/N$ entries. 
For the triangular RPFM the situation changes completely since here the 
average matrix $\langle G_{ij}\rangle=1/(j-1)$ (for $i<j$ and $j\ge 2$)
has already a non-trivial structure and eigenvalue spectrum. Therefore the 
argument of degenerate perturbation theory which allowed to apply the results 
of standard full non-symmetric random matrices does not apply here. 
In Fig.~\ref{fig16} one clearly sees that for $N=400$ the spectra for 
one realization of a triangular RPFM and its average are very similar 
for the eigenvalues with large modulus but both do not have at all a uniform 
circular density in contrast to the RPRM models without the triangular 
constraint discussed above. 
For the triangular RPFM the PageRank behaves as $P(K)\sim 1/K$ 
with the ranking index $K$ being close to the natural order of nodes 
$\{1,2,3,\ldots\}$ that reflects the fact that the node 1 has the maximum 
of $N-1$ incoming links etc.

The study of above models shows that it is not so simple to
find a good  RPFM model which reproduces a typical spectral structure
of real directed networks.

\section{Discussion}

In this study we presented a detailed analysis of the spectrum of the CNPR
for the period 1893 -- 2009.
It happens that the numerical simulations should be done
with a high accuracy (up to $p=16384$ binary digits for 
the rational interpolation method or $p=768$ binary digits for the 
high precision Arnoldi method)
to determine correctly the eigenvalues of the Google matrix
of CNPR at small eigenvalues $\lambda$.
Due to the time ordering of citations, 
the CNPR $G$ matrix  is close to the triangular  form
with a nearly nilpotent matrix structure. 
We show that special semi-analytical methods
allow to determine efficiently
the spectrum of such matrices. The eigenstates
with large modulus of $\lambda$ are shown to select
specific communities of articles in  certain 
research fields but there is no clear way
on how to identify a community
one is interested in. 

The obtained results
show that the spectrum of CNPR is characterized by the fractal Weyl law
with the fractal dimension $d_f \approx 1$ and the growth exponent
$b \approx 0.5$ being significantly smaller than unity. 
We think that the Phys. Rev. network has a structure
which is typical for other citation networks
and thus our result shows that the fractal Weyl law is 
a typical feature of citation networks.

The ranking of articles is analyzed with the help
of PageRank and CheiRank vectors corresponding to forward
and backward citation flows in time. It is shown that 
the correlations between these two vectors are small
and even negative that is similar to the case of
Linux Kernel networks \cite{2dmotor}
and significantly different from networks of universities
and Wikipedia. The 2DRanking on the PagRank-CheiRank plane
allows to select articles which efficiently redistribute information
flow on the CNPR.

To characterize the local impact propagation for a given article
we introduce the concept of ImpactRank which efficiently determines
its domain of influence.

Finally we perform the analysis of several  models of RPFM showing that
such full random matrices are very far from the realistic cases
of directed networks. Random sparse matrices with a limited number $Q$ of 
links per nodes seem to be closer to typical Google matrices concerning the 
matrix structure. 
However, still such random models give a rather uniform eigenvalue density 
with a spectral radius $\sim 1/\sqrt{Q}$ and also a flat PageRank 
distribution. Furthermore they do not 
capture the existence of quasi-isolated communities
which generates quasi-degenerate spectrum at $\lambda=1$.
Further development of RPFM models is required to 
reproduce the spectral properties of real modern directed networks.

In summary, we developed powerful numerical methods which allowed us to 
determine numerically the exact eigenvalues and eigenvectors of
the Google matrix of Physical Review. We demonstrated that this 
matrix is close to triangular matrices of large size where numerical errors
can significantly affect the eigenvalues. We show that the techniques 
developed in this work allow to resolve such difficulties
and obtain the exact spectrum in a semi-analytical manner.
The eigenvectors of eigenvalues with $|\lambda|<1$
are located on certain communities of articles
related to specific scientific research subjects. 
We point that the random matrix models of Google matrices are
still waiting their detailed development. Indeed, matrices with random elements 
have a spectrum being very different from the real one.
Thus, while the random matrix theory of Hermitian and unitary matrices
has been very successful (see e.g. \cite{mehta}), a random matrix theory
for Google matrices still waits its development. It is possible that 
the case of triangular matrices, which is rather similar to 
our CNPR case, can be a good starting point for development
of such models.

\section{Acknowledgments}
We thank the American Physical Society for letting us use their
citation database for Physical Review \cite{physrev_data}.
This research is supported in part by the EC FET Open project 
``New tools and algorithms for directed network analysis''
(NADINE $No$ 288956). This work was granted access to the HPC resources of 
CALMIP (Toulouse) under the allocation 2012-P0110.

\appendix

\section{Theory of triangular adjacency matrices}

Let us briefly remind the analytical theory of \cite{integers} 
for pure triangular networks with a nilpotent matrix $S_0$ such that 
$S_0^l=0$. For integers   \cite{integers} the adjacency
matrix is definded as   $A_{mn} = k$ where $k$ is a ‘multiplicity’ defined 
as the largest integer such that $m^k$ is a
divisor of $n$ and if $1 < m < n$, and $k = 0$ if $m = 1$ or 
$m = n$ or if $m$ is not a divisor of $n$. Thus,
we have $k = 0$ if $m$ is not a divisor of $n$ and $k \geq 1$ 
if $m$ is a divisor of $n$ different from $1$ and $n$.
The total size N of the matrix is fixed by the maximal considered integer.
Then the Google matrix is constructed from $A_{mn}$
following the standard rules described above.
This network of integers gives an important example of
a triangular Google matrix with a similar features also appearing
in the Physical Review citation network. Below we discuss the general
properties of such matrices.

For this we define the coefficients:
\begin{equation}
\label{eq_coeffs}
c_j=d^T S_0^j\,e/N\quad,\quad b_j=e^T S_0^j\,e/N
\end{equation}
which are non-zero only for $j=0,\,1,\,\ldots,\,l-1$. The fact that the 
non-vanishing columns of $S_0$ are sum normalized and that the other columns 
(corresponding to dangling nodes) are zero can be written as: 
$e^T S_0=e^T-d^T$ implying $d^T=e^T(\openone -S_0)$. Using this identify and 
the fact that $S_0^k=0$ for $k\ge l$ we find:
\begin{equation}
\label{eq_sum1}
%b_j=\sum_{k=j}^{l-1} c_k
\sum_{k=j}^{l-1} c_k=d^T(\openone-S_0)^{-1} S_0^j\, e/N=
e^T S_0^j\,e/N=b_j
\end{equation}
and in particular for $j=0$ we obtain the sum rule $\sum_{k=0}^{l-1} c_k=1$ 
and for $j=l-1$ the identity $b_{l-1}=c_{l-1}$. 

Consider now a right eigenvector $\psi$ of $S$ with eigenvalue $\lambda$. 
If $d^T\psi=0$ we find from (\ref{eq_matrixS}) that $\psi$ is also an 
eigenvector 
of $S_0$ and since $S_0$ is nilpotent the eigenvalue must be $\lambda=0$. 
Therefore for $\lambda\neq 0$ we have necessarily $d^T\psi\neq 0$ and 
with the appropriate normalization of $\psi$ we have $d^T\psi=1$ 
that implies together with the eigenvalue equation: 
$\psi=(\lambda\openone-S_0)^{-1}\,e/N$ where the matrix inverse is well 
defined for $\lambda\neq 0$. The eigenvalue is determined by the condition:
\begin{equation}
\label{eq_eig1}
0=\lambda^l(1-d^T\psi)=\lambda^l\left(1-d^T
\frac{\openone}{\lambda\openone-S_0}e/N
\right)\ .
\end{equation}
Since $S_0$ is nilpotent we may expand the matrix inverse in a finite series 
and therefore the eigenvalue $\lambda$ is the zero of the reduced polynomial
of degree $l$:
\begin{equation}
\label{eq_polyred}
{\cal P}_r(\lambda)=\lambda^l-\sum_{j=0}^{l-1}\lambda^{l-1-j}\,c_j
\end{equation}
where the coefficients $c_j$ are given by (\ref{eq_coeffs}). 
Using $d^T=e^T(\openone -S_0)$ 
we may rewrite (\ref{eq_eig1}) in the form:
\begin{equation}
\label{eq_eig2}
0=\lambda^l\left(1-e^T\frac{\openone-S_0}{\lambda\openone-S_0}e/N
\right)=(\lambda-1)\lambda^l\,e^T\frac{\openone}{\lambda\openone-S_0}e/N
\end{equation}
which gives another expression for the reduced polynomial:
\begin{equation}
\label{eq_polyred2}
{\cal P}_r(\lambda)=(\lambda-1)\sum_{j=0}^{l-1}\lambda^{l-1-j}\,b_j
\end{equation}
using the coefficients $b_j$ and confirming explicitly that $\lambda=1$ 
is indeed an eigenvalue of $S$. The expression (\ref{eq_polyred2}) 
can also be obtained by a direct calculation from (\ref{eq_sum1}) 
and (\ref{eq_polyred}). 

Since the reduced polynomial has at most $l$ zeros $\lambda_j$ 
($\neq 0$ since $c_{l-1}=b_{l-1}\neq 0$) we find that there are at most 
$l$ non-vanishing eigenvalues of $S$ given by these zeros. They can also 
be obtained as the eigenvalues of a ``small'' $l\times l$ matrix. To see 
this let us define the following set of vectors $v_j$ for $j=1,\,\ldots,\,l$ by
$v_j=c_{j-1}^{-1}\,S_0^{j-1}\,e/N$ where we have chosen to apply 
the prefactor $c_{j-1}^{-1}$ to the vector $S_0^{j-1}\,e/N$ 
\cite{discuss_prefactor}. From (\ref{eq_matrixS}) and (\ref{eq_coeffs}) 
one finds that $Sv_j$ can be expanded in the other vectors $v_k$ as
\begin{equation}
\label{eq_Sapplied}
Sv_j=\frac{c_j}{c_{j-1}}\,v_{j+1}+c_0\,v_1=\sum_{k=1}^l{\bar S}_{kj}\,v_k
\end{equation}
where ${\bar S}_{kj}$ are the matrix elements of the $l\times l$ 
representation matrix 
\begin{equation}
\label{eq_repmatrix}
\bar S=
\left(\begin{array}{ccccc}
c_0 & c_0 & \cdots & c_0 & c_0 \\
c_1/c_0   & 0   & \cdots & 0 & 0 \\
0   & c_2/c_1   & \cdots & 0 & 0 \\
\vdots & \vdots & \ddots & \vdots & \vdots \\
0   & 0   & \cdots & c_{l-1}/c_{l-2} & 0 \\
\end{array}\right)\quad.
\end{equation}
Note that for the last vector $v_l$ we have $Sv_l=c_0\,v_1$ since 
$c_{l}=0$ and therefore the matrix ${\bar S}$ provides a closed and 
mathematically exact representation of $S$ on the $l$-dimensional subspace 
generated by $v_1,\,\ldots,\,v_l$. Furthermore one can easily verify (by 
a recursive calculation in $l$) that the 
characteristic polynomial of ${\bar S}$ coincides with the reduced polynomial 
(\ref{eq_polyred}). Therefore numerical diagonalization of ${\bar S}$ 
provides an alternative method to compute the non-vanishing eigenvalues of 
$S$. In 
principle one can also determine directly the zeros of the reduced polynomial 
by the Newton-Maehly method and in \cite{integers} this was indeed done 
for cases with very modest values of $l\le 29$. However, here for the 
triangular CNPR we have $l=352$ and 
the coefficients $c_j$ become very small, especially: 
$c_{l-1}\approx 3.6\times 10^{-352}$ a number which is (due to the exponent) 
outside the range of 64 bit standard double-precision numbers 
(IEEE 754) with 52 bits 
for the mantissa, 10 bits for the exponent (with respect to 2) and 
two bits for the signs of mantissa and exponent. This exponent range 
problem is not really serious and can for example be circumvented by a 
smart reformulation of the algorithm to evaluate 
the ratio ${\cal P}_r(\lambda)/{\cal P}_r'(\lambda)$ using only ratios 
$c_{j}/c_{j-1}$ which do not have this exponent range problem. However, 
it turns out that in this approach 
the convergence of the Newton-Maehly method using double-precision 
arithmetic is very bad for many zeros and does not provide 
reliable results. Below in Appendix B 
we show how this problem can be solved using 
high precision calculations but for the moment we mention that 
one may also try another approach by diagonalizing numerically the 
representation matrix ${\bar S}$ given in (\ref{eq_repmatrix}) which also 
depends on the ratios $c_{j}/c_{j-1}$. 

\section{Effects of numerical errors and high precision computations}

We remind that the Arnoldi method determines an orthonormal set of vectors 
$\zeta_1,\,\zeta_2,\,\zeta_3,\,\ldots,\,\zeta_{n_A}$ 
where the first vector $\zeta_1$ is obtained by normalizing a given 
initial vector and $\zeta_{j+1}$ is obtained 
by orthonormalizing $S\zeta_j$ to the previous vectors determined so far. 
It is obvious due to (\ref{eq_Sapplied}) that for the initial uniform 
vector $e$ each 
$\zeta_j$ is given by a linear combination of 
the vectors $v_k$ with $k=1,\,\ldots,\,j$. Since the subspace 
of $v_k$ for $k=1,\,\ldots,\,l$ is closed with respect to applications 
of $S$ the Arnoldi method 
should, in theory, break off at $n_A=l$ with a zero coupling element. 
The latter is given as the norm of $S\zeta_l$ othogonalized to 
$\zeta_1,\,\ldots,\,\zeta_{l}$ and if this norm vanishes the vector 
$\zeta_{l+1}$ cannot be constructed and the Arnoldi method has completely 
explored an $S$-invariant subspace of dimension $l$.

However, due to a strong effect of round-off errors and the fact that the 
vectors $v_j$ are numerically ``nearly'' linearly dependent 
the last coupling element does  not vanish numerically (when 
using double-precision) and the Arnoldi method produces a cloud 
of numerically incorrect eigenvalues due to the Jordan blocks
which are mathematically outside the representation space 
(defined by the vectors 
$v_j$) but which are still explored due to round-off errors and 
clearly visible in Fig.~\ref{fig5}. The double-precision spectrum of 
${\bar S}$ seems to provide well defined eigenvalues in the range where 
the Arnoldi method produces the ``Jordan block cloud'' but outside this cloud 
both spectra coincide  only partly, mainly for the eigenvalues 
with largest modulus and positive real part. For the eigenvalues with negative 
real part there are considerable deviations. As can be seen 
in Fig.~\ref{fig6} the 
eigenvalues produced by the Arnoldi method at double-precision are reliable 
provided that they are well {\em outside} 
the Jordan block cloud of incorrect eigenvalues. 
Therefore the deviations outside the Jordan block cloud show that 
the numerical double-precision diagonalization of the 
representation matrix 
${\bar S}$ is not reliable as well but here the effect of numerical 
errors is quite different as for the Arnoldi method as it is
explained below.

In order to obtain an alternative and reliable numerical method to determine 
the spectrum of the triagonal CNRP we have also 
tried to determine the zeros of the reduced polynomial using 
higher precision numbers with 80 or even 128 bits (quadruple precision) 
which helps to solve the (minor) exponent 
range problem (mentionned in Appendix A) 
because these formats use more bits for the exponent. 
However, there are indeed two other serious numerical problems. 
First it turns out that in a certain range of 
the complex plane around $\mbox{Re}(\lambda)\approx-0.1$ to $-0.2$ 
and 
$\mbox{Im}(\lambda)\le 0.1$ the numerical evaluation of the polynomial suffers 
in a severe way from an alternate sign problem with a strong 
loss of significance. 
Second the zeros of the polynomial depend in a very sensitive way on 
the precision of the coefficients $c_j$ (see below). 
We have found that even 128 bit numbers 
are not sufficient to obtain all zeros with a reasonable graphical precision. 

Therefore we use the very efficient 
GNU Multiple Precision Arithmetic Library (GMP library) 
\cite{gmplib}. With this library one has 31 bits for the exponent and 
one may chose an arbitrary number of bits for the mantissa. 
We find that using 256 bits (binary digits) for the mantissa 
the complex zeros of the reduced polynomial can be determined with a 
precision of $10^{-18}$. In this case the convergence of the Newton-Maehly 
method is very nice and we find that the sum (and product) of the 
complex zeros coincide with a high precision with the theoretical values 
$c_0$ (respectively: $(-1)^{l-1}c_{l-1}$) due to (\ref{eq_polyred}). 
We have also tested 
different ways to evaluate the polynomial, such as Horner scheme versus 
direct evaluation of the sum and for both methods using both expressions 
(\ref{eq_polyred}) and (\ref{eq_polyred2}). It turns out that 
with 256 binary digits during the calculation the zeros obtained by 
the different variants of the method coincide very well within the 
required precision of $10^{-18}$. Of course the coefficients $c_j$ or 
$b_j$ given by (\ref{eq_coeffs}) need also to be evaluated with the 
precision of 256 binary digits but there is no problem of using high 
precision vectors since the non-vanishing matrix 
elements of $S_0$ are 
rational numbers that allow to perform the evaluation of the 
vectors $S_0^j e/N$ with arbitrary precision. 
We also tested a random modification of $c_j$ according to 
$c_j\to c_j(1+10^{-16}X)$ where $X$ is a random number in the 
interval $]-0.5,\,0.5[$. This modification gives significant differences 
of the order of $10^{-2}$ to $10^{-1}$ for some of the 
complex zeros and which 
are well visible in the graphical representation of the spectra. 
Therefore, the spectrum depends in a very sensitive way on these 
coefficients and it is now quite clear that numerical double-precision 
diagonalization of ${\bar S}$, which depends according 
to (\ref{eq_repmatrix}) on the values $c_j$, 
cannot provide accurate eigenvalues simply because the double-precision 
round-off errors of $c_j$ imply a sensitive change of eigenvalues. 
In particular some of the numerical eigenvalues of ${\bar S}$ differ 
quite strongly from the high precision zeros of the reduced polynomial. 

In order to study more precisely the effect of the numerical instability 
of the Arnoldi method due to the Jordan blocks we also use the 
GMP library to increase the numerical precision of the Arnoldi method. 
To be precise we implement the first part of this method, 
the {\em Arnold iteration} in which the $n_A\times n_A$ Arnoldi 
representation matrix is determined by the Gram-Schmidt orthogonalization 
procedure, using high precision numbers while for the second 
step, the numerical diagonalization of this representation matrix, 
we keep the standard double-precision. 
It turns that only the first step is numerically critical. Once the 
Arnoldi representation matrix is obtained in a careful and precise way, 
it is numerically well conditioned and its numerical diagonalization 
works well with only double-precision.

\section{Theory of degenerate eigenvalues}

In order to understand the mechanism of the degenerate core space 
eigenvalues visible in Fig.~\ref{fig8} 
we extend the argumentation of Appendix A 
for triangular CNPR to the case of nearly triangular networks. 
Consider again the matrix $S$ given by Eq. (\ref{eq_matrixS}) but now 
$S_0$ is not nilpotent. 
There are two groups of eigenvectors $\psi$ of $S$ with eigenvalue $\lambda$. 
The first group is characterized by the orthogonality $d^T\psi=0$ of the 
eigenvector $\psi$ with respect to the dangling vector $d$ and the 
second group is characterized by the non-orthogonality $d^T\psi\neq 0$. 
In the following, we describe efficient methods to determine all 
eigenvalues of the first group and 
a considerable number of eigenvalues
of the second group. 
We note that for the case of a purely triangular network the first group 
contains only eigenvectors for the eigenvalue 0 and the second group contains 
the eigenvectors for the $l$ non-vanishing eigenvalues as discussed in 
Appendix A. 
In principle there are also complications due to generalized eigenvectors 
(associated to non-trivial Jordan blocks) but they appear mainly for the 
eigenvalue zero and for the moment we do not discuss these complications.

First we note that the subspace eigenvectors of $S$ belong to the first 
group because the nodes of the subspaces of $S$ cannot contain 
dangling nodes which, by construction of $S$, 
are linked to any other node and therefore belong to the core space. 
Since any subspace eigenvector $\psi$ has non-vanishing values only for 
subspace nodes being different from dangling nodes 
we have obviously $d^T\psi=0$. 
We also note that an eigenvector of $S$ of the first group with $d^T\psi=0$ 
is due to (\ref{eq_matrixS}) also an eigenvector of $S_0$ with the 
same eigenvalue.

For the remaining eigenvectors in the first group one might try 
to diagonalize the matrix $S_0$ and check for each eigenvector of $S_0$ if the 
identity $d^T\psi=0$ holds 
in which case we would obtain an eigenvector of $S$ of 
the first group but generically, and apart from the subspace 
eigenvectors, there is no reason that eigenvectors of 
$S_0$ with isolated non-degenerate eigenvalues obey this identity. 
However, if we have an eigenvalue of $S_0$ with a degeneracy $m\ge 2$ 
we may construct by suitable linear combinations $m-1$ linearly independent 
eigenvectors of $S_0$ which also obey $d^T\psi=0$ and therefore 
this eigenvalue with degeneracy $m$ of $S_0$ is also an eigenvalue 
with degeneracy $m-1$ of $S$. In order to determine the degenerate 
eigenvalues of $S_0$ it is useful to determine the subspaces of $S_0$ which 
(in contrast to the subspaces of $S$) may contain dangling nodes. Actually, 
each dangling node is a trivial invariant subspace (for $S_0$) of 
dimension 1 with a network 
matrix of size $1\times 1$ and being zero. 
Explicitly we have implemented the following procedure: first we determine 
the subspaces of $S$ (with 71 nodes in total) and remove these nodes from the 
network. Then we determine all subspaces of $S_0$ whose dimension is below 10. 
Each time such a subspace is found its nodes are 
immediately removed from the network. When we have tested in a first run 
all nodes as potential subspace nodes the procedure is repeated until 
no new subspaces of maximal dimension 10 are found 
since removal of former subspaces may have created new subspaces. 
Then the limit size of 10 is doubled to 20, 40, 80 etc. to ensure that we do 
not miss large subspaces. However, for the CNPR it 
turns out that the limit size of 10 allows to find all subspaces. 
In our procedure a subsequently found 
subspace may potentially have links to a former subspace leading to a 
block-triangular (and not block-diagonal structure as it was done 
in ref. \cite{univuk}). This method to determine ``relative'' subspaces 
of a network already reduced by former subspaces is more convenient for the 
CNPR which is nearly triangular and it allows also to determine 
correctly all subspace eigenvalues by diagonalizing each relative subspace 
network. The removal of subspace nodes of $S$ and $S_0$ reduces the network 
size from $N=463348$ to $404959$. 
In the next step we remove in the same way the subspaces of the transpose 
$S_0^T$ 
of $S_0$ (since the eigenvalues of $S_0^T$ and $S_0$ are the same) which 
reduces the network size furthermore to $90965$. 
In total this procedure provides a block triangular structure of $S_0$ 
as:
\begin{equation}
\label{eq_S0block}
S_0=
%\left(\begin{array}{ccccccc}
\left(\begin{array}{ccccccc}
S_1 & * & \cdots && \phantom{\vdots} & \cdots & * \\
0 & S_2 & * &&&& \vdots \\
\vdots & \ddots & \ddots & \ddots & \phantom{\ddots} & \phantom{\ddots} & \vdots \\
\phantom{\ddots}  && 0 & B & * && \phantom{\vdots}  \\
\vdots  &&& 0 & T_1 & * & \vdots  \\
\vdots &&&& 0 & T_2 & * \\
0 &\cdots &&& \cdots & \ddots & \ddots \\
\end{array}\right)
\end{equation}
where $S_1,\,S_2,\,\ldots$ represent the diagonal subblocks associated 
to the subspaces of $S$ and $S_0$ while $T_1,\,T_2,\,\ldots$ represent the 
diagonal subblocks associated to the subspaces of $S_0^T$ and 
$B$ is the ``bulk'' part for the remaining network of $90965$ nodes. 
The stars represent potential non-vanishing entries whose values do not 
influence the eigenvalues of $S_0$. 
The subspace blocks $S_1,\,S_2,\,\ldots$ and $T_1,\,T_2,\,\ldots$ 
which are individually of maximal dimension 10 can be directly diagonalized 
and it turns that out of $372382$ eigenvalues in these blocks only 
about $4000$ eigenvalues (counting degeneracies) or $950$ eigenvalues 
(non-counting degeneracies) are different from zero. Most of these eigenvalues 
are not degenerate and are therefore not eigenvalues of $S$ but 
there are still quite many degenerate eigenvalues at $\lambda=\pm 1/\sqrt{n}$ 
with $n\ge 2$ taking small integer values and who are also eigenvalues 
of $S$ with a degeneracy reduced by one.

Concerning the bulk block $B$ we can write it in the form 
$B=B_0+f_1\,e_1^T$ 
%%where $f_1\,e_1^T$ is the contribution of the first column of 
where $f_1$ is the first column vector of $B$ and 
$e_1^T=(1,\,0,\,\ldots,\,0)$. 
The matrix $B_0$ is obtained from $B$ by replacing its first column 
to zero. We can apply the above argumentation between $S$ and $S_0$ 
in the same way to $B$ and $B_0$, i.e. the degenerate eigenvalues of $B_0$ 
with degeneracy $m$ are also eigenvalues of $B$ with degeneracy $m-1$ (with 
eigenvectors obeying $e_1^T \psi=0$) 
and therefore eigenvalues of $S$ with degeneracy $m-2$. The matrix $B_0$ is 
decomposed in a similar way as in (\ref{eq_S0block}) 
with subspace blocks, which can be diagonalized numerically, 
and a new bulk block $\tilde B$ of dimension $63559$ and which may 
be treated in the same way by taking out its first column. 
This procedure provides a recursive scheme which after 9 iterations 
stops with a final bulk block of zero size. At each iteration we keep 
only subspace eigenvalues with degeneracies $m\ge 2$ and which are joined 
with reduced degeneracies $m-1$ to the subspace spectrum of the 
previous iteration. For this joined spectrum we keep again 
only eigenvalues with degeneracies $m\ge 2$ which are joined with 
the subspace spectrum of the next higher level etc. 

In this way we have determined all eigenvalues of $S_0$ with a 
degeneracy $m\ge 2$ which belong to the eigenvalues of $S$ of the first group. 
Including the direct subspace of $S$ there are 
4999 non-vanishing eigenvalues (counting 
degeneracies) or 442 non-vanishing eigenvalues (non-counting 
degeneracies). The degeneracy of the zero eigenvalue (or the dimension 
of the generalized kernel) is found by this procedure to be 
$455789$ but this would only be correct assuming that there are no general 
eigenvectors of higher order (representation vectors of non-trivial Jordan 
blocks) which is clearly not the case. 
The Jordan subspace structure of the zero eigenvalue complicates the 
argumentation. Here at each iteration step the degeneracy has to be reduced 
from $m$ to $m-D$ where $D>1$ is the dimension of the maximal Jordan block 
since each generalized eigenvector at a given order has to be treated as an 
independent vector when constructing vectors obeying the orthogonality 
with respect to the dangling vector $d$. 
Therefore the degeneracy of the zero eigenvalue cannot be determined 
exactly but we may estimate its degeneracy 
of about $\sim 455000$ out of $463348$ nodes in total. This implies that the 
number of non-vanishing eigenvalues is about $\sim 8000-9000$ which is 
considerably larger than the value of $352$ for the triangular 
CNPR but still much smaller than the total 
network size. 

%%% table for degeneracies !!!
\begin{table}
\centering
\include{table_degen}
\caption{\label{table1}
Degeneracies of the eigenvalues with largest modulus for the 
whole CNPR whose eigenvectors $\psi$ belong to the first 
group and obey the orthogonality $d^T\psi=0$ with the dangling vector $d$. }
\end{table}

In Table~\ref{table1} we provide the degeneracies for 
some of the eigenvalues $\pm 1/\sqrt{n}$ for integer $n$ in the range 
$1\le n\le 25$. The degeneracies for $+1/\sqrt{n}$ and $-1/\sqrt{n}$ 
are identical for non-square numbers $n$ (with non-integer $\sqrt{n}$) 
and different for square numbers (with integer $\sqrt{n}$). 
Apparently for non-square numbers the eigenvalues are only generated from 
effective $2\times 2$ blocks:
\begin{equation}
\label{eq_2x2blocks}
\left(\begin{array}{cc}
0 & 1/n_1  \\
1/n_2 & 0  \\
\end{array}\right)
\quad\Rightarrow\quad \lambda=\pm \frac{1}{\sqrt{n_1\,n_2}}
\end{equation}
with positive integers $n_1$ and $n_2$ such that $n=n_1\,n_2$ 
while for square numbers $n=m^2$ they 
may be generated by such blocks or by simple $1\times 1$ blocks containing 
$1/m$ such that the 
degeneracy for $+1/\sqrt{n}=+1/m$ is larger than the degeneracy for 
$-1/\sqrt{n}=-1/m$. 
Furthermore, statistically the degeneracy is smaller for prime numbers $n$ or 
numbers with less factorization possibilities and larger for numbers with more 
factorization possibilities. 
The Arnoldi method (with 52 binary digits for double-precision arithmetic and
$n_A=8000$) provides according to the sizes of the plateaux 
visible in Fig.~\ref{fig8} 
the overall approximate degeneracies $\sim 60$ for $|\lambda|=1/\sqrt{2}$ 
(i.e. $\pm 1/\sqrt{2}$ counted together), $\sim 50$ for 
$|\lambda|=1/\sqrt{3}$ and $\sim 115$ for $|\lambda|=1/2$. These values 
are coherent with (but slightly larger than) the values $54$, $40$ and 
$110$ taken from Table~\ref{table1}. Actually, as we will see below, the 
slight differences between the degeneracies obtained from Fig.~\ref{fig8} 
and from Table~\ref{table1} are indeed relevant and correspond to some 
eigenvalues of the second group which are close but not identical 
to $\pm 1/\sqrt{2}$, $\pm 1/\sqrt{3}$ or $\pm 1/2$ and do not contribute 
in Table~\ref{table1}.

\section{Rational interpolation method}

We now consider the eigenvalues $\lambda$ of $S$ 
for the eigenvectors of the second 
group with non-orthogonality $d^T\psi\neq 1$ or $d^T\psi=1$ after 
proper renormalization of $\psi$. Now $\psi$ cannot be an eigenvector 
of $S_0$ and $\lambda$ is not an eigenvalue of $S_0$. Similarly as in 
Appendix A the eigenvalue equation $S\psi=\lambda\psi$, the condition 
$d^T\psi=1$ and (\ref{eq_matrixS}) 
imply that the eigenvalue $\lambda$ of $S$ is a zero of the rational 
function
\begin{equation}
\label{eq_rationalfunction}
{\cal R}(\lambda)=1-d^T
\frac{\openone}{\lambda\openone-S_0}e/N=1-\sum_{j,q} 
\frac{C_{jq}}{(\lambda-\rho_j)^q}
\end{equation}
where we have formally expanded the vector $e/N$ in eigenvectors of $S_0$ and 
with $\rho_j$ being the eigenvalues of $S_0$ and $q$ is 
the {\em order} of the eigenvector of $\rho_j$ used in this expansion, 
i.e. $q=1$ for simple eigenvectors and $q>1$ for generalized eigenvectors of 
higher order due to Jordan blocks. Note that even the largest possible 
value of $q$ for a given eigenvalue may be (much) smaller than its 
multiplicity $m$. Furthermore the case of simple repeating 
eigenvalues (with simple eigenvectors) with higher multiplicity $m>1$ 
leads only to several identical terms $\sim (\lambda-\rho_j)^{-1}$ 
for any eigenvector of this eigenvalue thus all contributing to the 
coefficients $C_{jq}$ and whose precise values we do not need to know 
in the following. For us the important point is that the second 
identity in (\ref{eq_rationalfunction}) establishes that ${\cal R}(\lambda)$ 
is indeed a rational function whose denominator and numerator polynomials
have the same degree and whose poles are (some of) the eigenvalues of $S_0$. 

We mention that one can also show by a simple determinant calculation 
(similar to a calculation shown in \cite{integers} for triangular 
networks with nilpotent $S_0$) that:
\begin{equation}
\label{eq_characpoly}
P_S(\lambda)=P_{S_0}(\lambda)\,{\cal R}(\lambda)
\end{equation}
where $P_S(\lambda)$ [or $P_{S_0}(\lambda)$] is the characteristic polynomial 
of $S$ ($S_0$). Therefore those zeros of ${\cal R}(\lambda)$ 
which are not zeros of $P_{S_0}(\lambda)$ (i.e. not eigenvalues of $S_0$)
are indeed zeros of $P_S(\lambda)$ (i.e. eigenvalues of $S$) since there are 
not poles of ${\cal R}(\lambda)$. 
Furthermore, generically the {\em simple} zeros $P_{S_0}(\lambda)$ also 
appear as poles in ${\cal R}(\lambda)$ and are therefore not 
eigenvalues of $S$. 
However, for a zero of $P_{S_0}(\lambda)$ (eigenvalue of $S_0$) 
with {\em higher multiplicity} $m>1$ (and unless $m$ is equal to the maximal 
Jordan block order $q$ associated to this eigenvalue of $S_0$) 
the corresponding pole in ${\cal R}(\lambda)$ only 
reduces the multiplicity to $m-1$ (or $m-q$ in case of higher order 
generalized eigenvectors) and we have also a zero of $P_S(\lambda)$ 
(eigenvalue of $S$). Some of the eigenvalues of $S_0$, whose eigenvectors 
$\psi$ are orthogonal to the dangling vector ($d^T\psi=0$) 
and do not contribute in the expansion in (\ref{eq_rationalfunction}), 
are not poles of 
${\cal R}(\lambda)$ and therefore also eigenvalues of $S$. This concerns 
essentially the direct subspace eigenvalues of $S$ which are also 
direct subspace eigenvalues of $S_0$ as already discussed in Appendix C. In 
total the identity (\ref{eq_characpoly}) confirms exactly the above picture 
that there are two groups of eigenvalues and with the special role 
of direct subspace eigenvalues belonging to the first group. 

Our aim is to determine numerically the zeros of the rational function 
${\cal R}(\lambda)$. In order to evaluate this function 
we expand the first identity in (\ref{eq_rationalfunction})
in a matrix geometric series and we obtain
\begin{equation}
\label{eq_rationseries}
{\cal R}(\lambda)=1-\sum_{j=0}^\infty c_j \lambda^{-1-j}
\end{equation}
with the coefficients $c_j$ defined in (\ref{eq_coeffs}) and 
provided that this series converges. 
In Appendix A, where we discussed the case of a nilpotent matrix 
$S_0$ with $S_0^l=0$, the series was finite and for this particular case we 
had ${\cal R}(\lambda)=\lambda^{-l}{\cal P}_r(\lambda)$ 
where ${\cal P}_r(\lambda)$ was the reduced polynomial defined in 
(\ref{eq_polyred}) and whose zeros provided the $l$ non-vanishing eigenvalues 
of $S$ for nilpotent $S_0$. 

However, for the CNPR the series are infinite since 
all $c_j$ are different from zero. One may first try a crude approximation 
and simply replace the series by a finite sum for $j<l$ and using 
some rather large 
cutoff value for $l$ and determine the zeros in the same way 
as for the nilpotent case 
(high precision calculation of the zeros of the reduced polynomial of 
degree $l$). 
It turns that in this way we obtain correctly the largest 
core space eigenvalue of 
$S$ as $\lambda_1=0.999751822283878$ which is also obtained 
by (any variant of) the Arnoldi method. However, the other zeros obtained 
by this approximation lie all on a circle of radius $\approx 0.9$ in the 
complex plane and obviously do not represent any valid eigenvalues. 
Increasing the cutoff value $l$ does not help either and it 
increases only the density of zeros on this circle. 
To understand this behavior we note that in the limit $j\to\infty$ 
the coefficients $c_j$ behave as $c_j\propto \rho_1^j$ where 
%%%$\rho_1=0.90244828051922380056$ 
$\rho_1=0.902448280519224$ is the largest eigenvalue of the 
matrix $S_0$ with an eigenvector non-orthogonal to $d$. Note that 
the matrix $S_0$ has also some degenerate eigenvalues at $+1$ and $-1$ but 
these eigenvalues are obtained from the direct subspace eigenvectors of $S$ 
(which are also direct subspace eigenvectors of $S_0$) and which are 
orthogonal to the dangling vector $d$ and do not contribute in 
the rational function (\ref{eq_rationalfunction}). It turns actually out 
that the eigenvalue $\rho_1$ is also the largest {\em subspace space}
eigenvalue of $S_0$ (after having removed the direct subspace nodes of $S$). 
By analyzing explicitly the small-dimensional 
subspace related to this eigenvalue one can 
show that $\rho_1$ is given as the largest solution of the polynomial equation 
$x^3-\frac{2}{3}x-\frac{2}{15}=0$ and can therefore be expressed as 
$\rho_1=2\,\mbox{Re}\,[(9+i\sqrt{119})^{1/3}]/(135)^{1/3}$. The asymptotic 
behavior $c_j\propto \rho_1^j$ is also confirmed by the direct numerical 
evaluation of $c_j$. Therefore the series (\ref{eq_rationseries}) 
converges only for $|\lambda|>\rho_1$ and a simple (even very large) 
cutoff in the sum implies that only 
eigenvalues $|\lambda_j|>\rho_1$ can be determined as a zero of the finite 
sum. The only eigenvalue respecting this condition is the largest core space 
eigenvalue $\lambda_1$ given above.

One may try to improve this by a ``better'' approximation which consists of 
evaluating the sum exactly up to some value $l$ and than to replace 
the remaining sum as a geometric series with the approximation: 
$c_j\approx c_l\rho_1^{j-l}$ for $j\ge l$ and with $\rho_1$ determined 
as the ratio $\rho_1=c_{l}/c_{l-1}$ (which provides a sufficient 
approximation) or taken as its exact (high precision) value. 
This improved approximation 
results in ${\cal R}(\lambda)\approx \lambda^{-l}
(\lambda-\rho_1)^{-1} {\cal P}(\lambda)$ with a polynomial ${\cal P}(\lambda)$
whose zeros provide in total four correct eigenvalues. Apart from 
$\lambda_1$ it also gives $\lambda_2=0.902445536212661$ (note that this 
eigenvalue of $S$ is very close but different to the 
eigenvalue $\rho_1$ of $S_0$) 
and $\lambda_{3,4}=0.765857950563684\pm i\,0.251337495625571$ 
such that $|\lambda_{3,4}|=0.806045245100386$. All these four core space 
eigenvalues coincide very well with the first four eigenvalues obtained 
from the Arnoldi method. 
However, the other zeros of the Polynomial ${\cal P}(\lambda)$ 
lie again on a circle, now with a reduced radius $\approx 0.7$, and do not 
coincide with eigenvalues of $S$. 
This can be understood by the fact that the coefficients $c_j$ obey for 
$j\to\infty$ the more precise asymptotic expression 
$c_j\approx C_1\rho_1^j+C_2\rho_2^j+C_2\rho_3^j+\ldots$
with the next eigenvalues $\rho_2=1/\sqrt{2}\approx 0.707$ 
and $\rho_3=-\rho_2$. Here the first term $C_1\rho_1^j$ is
dealt with analytically by the replacement of the geometric series but 
the other terms create a new convergence problem. 
Therefore the improved approximation allows only to 
determine the four core space eigenvalues with 
$|\lambda_j|>|\rho_{2,3}|=1/\sqrt{2}$. 
To obtain more valid eigenvalues it seems to be necessary 
to sum up by geometric series many of the next terms, not only the 
next two terms due to $\rho_2$ and $\rho_3$, but also the following terms of 
smaller eigenvalues $\rho_j$ of $S_0$. In other words the 
exact pole structure of the rational function ${\cal R}(\lambda)$ has be kept
as best as possible.

Therefore due to the rational structure of the function ${\cal R}(\lambda)$ 
with many eigenvalues $\rho_j$ of $S_0$ that determine its precise pole 
structure 
we suggest the following numerical approach using high precision 
arithmetic. For a given number $p$ of binary digits, e.g. $p=1024$, 
we determine the coefficients $c_j$ for $j<l$ where 
the cutoff value 
\begin{equation}
\label{eq_lcutoff}
l\approx \frac{\ln(1-\rho_1)-p\ln(2)}{\ln(\rho_1)}
\approx 6.753\, p+{\rm const.}
\end{equation}
%$l\approx [\ln(1-\rho_1)-p\ln(2)]/\ln(\rho_1)$ 
is sufficiently 
large to evaluate the sum (\ref{eq_rationseries}) accurately in the 
given precision of $p$ binary digits (error below $2^{-p}$) 
for {\em all complex values $\lambda$ on the unit circle}, 
i.e. $|\lambda|=1$, where the series converges well. 
Furthermore we choose a number $n_R$ of ``eigenvalues'' we want to calculate, 
e.g. $n_R=300$, and evaluate the rational function 
${\cal R}(z)$ at $n_S=2 n_R+1$ support points $z_j=\exp(2\pi i\,j/n_S)$ 
($j=0,\,\ldots,\,n_S-1$) uniformly distributed on the unit circle 
using the series (\ref{eq_rationseries}). Then we calculate 
the rational function $R_I(z)$ which interpolates ${\cal R}(z)$ at the 
$n_S$ support points $z_j$, $R_I(z_j)={\cal R}(z_j)$, 
using Thiele's interpolation formula. Then 
the numerator and denominator  polynomials of $R_I(z)$ are both of 
degree $n_R$. 
Thiele's interpolation formula expresses $R_I(z)$ in terms of a continued 
fraction expansion using inverse differences. This method is quite 
standard and well described in the literature of numerical mathematics, 
see for example \cite{stoer}. 
After having evaluated a table of $n_S$ inverse differences 
(with $n_S^2/2$ operations) one can evaluate arbitrary values 
of $R_I(z)$ using the continued fraction expansion (with $n_S$ operations). 
It is not very difficult to derive from the continued fraction expansion 
a recursive scheme to evaluate the values of the numerator and denominator 
polynomials separately as well as their derivatives. 
Using this scheme we determine 
the $n_R$ complex zeros of the numerator polynomial using the 
(high precision variant of the) Newton-Maehly method. These zeros correspond 
to the zeros of the rational functional ${\cal R}(z)$ and are 
taken as approximate eigenvalues of the matrix $S$ of the second group. The 
main idea of this approach is to evaluate these zeros from the analytical 
continuation of ${\cal R}(z)$ using values for $|z|=1$ to determine 
its zeros well inside the unit circle. 

We also consider a second variant of the method where the number of 
support points $n_S=2 n_R+2$ is even (instead of $n_S=2 n_R+1$ being 
odd as for the first variant). In this case the numerator 
polynomial is of degree $n_R+1$ (instead of $n_R$) 
while the denominator polynomial 
is of degree $n_R$ and we choose to interpolate the inverse of the 
rational function $1/{\cal R}(z)$ (instead of ${\cal R}(z)$ itself) 
by $R_I(z)$ such that the zeros 
of ${\cal R}(z)$ are given by the $n_R$ zeros of the denominator (instead 
of the numerator) polynomial of $R_I(z)$. 

The number $n_R$ must not be too small 
in order to well approximate the second identity in 
(\ref{eq_rationalfunction}) by the fit function. On the other hand for a given 
precision of $p$ binary digits the number of $n_R$ must not be too large as 
well because the coefficients $c_j$, which may be written as the expansion 
$c_j=\sum_\nu C_\nu\,\rho_\nu^j$, do not contain enough information 
to resolve its structure for the smaller eigenvalues $\rho_j$ of 
$S_0$. Therefore for too large values of $n_R$ (for a given precision), 
we obtain additional artificial zeros of the numerator polynomial 
(or of the denominator polynomial for the second variant) of $R_I(z)$, mostly 
close to the unit circle, somehow as additional nodes around the 
support points. 

It turns out that for the proper combination of $p$ and $n_R$ values the 
method provides highly accurate eigenvalues and works astonishingly well. 
In particular for values of $n_R$ below a certain threshold (depending on 
the precision $p$) both variants of the method with odd or even number 
of support points provide numerically identical zeros (with 
final results rounded to 52 binary digits) which indeed 
coincide very accurately (for most of 
them) with the eigenvalues of $S$ we want to determine.

We remind that the rational interpolation method allows only to determine 
the eigenvalues of $S$ of the second group, i.e. the eigenvalues which are 
not eigenvalues of $S_0$ and whose eigenvectors obey $d^T\psi\neq 0$. 
The eigenvalues of the first group (with $d^T\psi= 0$) have to be
determined separately by the scheme of degenerate subspace 
eigenvalues of $S_0$ described in Appendix C. In particular the eigenvalues given 
in Table~\ref{table1} and belonging to the first group are not zeros of the 
rational function ${\cal R}(z)$ (they are actually poles of this function) 
but it turns out that there are some zeros 
of ${\cal R}(z)$ which are very close but not identical to some of the 
values in Table~\ref{table1}. For example the rational interpolation 
method provides the following zeros: $1/2+3.13401098\times 10^{-5}$,  
$1/2+1.3279300\times 10^{-7}$, $1/\sqrt{2}-1.1597\times 10^{-10}$
or $1/\sqrt{2}-6.419004\times 10^{-8}$ which are indeed accurate 
in the given precision since they are stable for all values 
of $p\ge 1024$ and the corresponding maximal value of $n_R$ 
and we have stopped the Newton iteration when 
the error of a zero was clearly below $10^{-18}$. 
These zeros are also found with the same precision 
in the data of the high precision Arnoldi method for the three 
different values of $256$, $512$ or $768$ binary digits. However, 
based only on results of the Arnoldi method it is not really clear if the 
small corrections to $1/2$ or $1/\sqrt{2}$ are real and exact or 
numerically artificial since the Arnoldi method has indeed problems 
with degenerate and clustered eigenvalues \cite{stewart}. 
Therefore the rational interpolation method provides an independent 
and strong confirmation 
of the accuracy of these type of eigenvalues. We attribute their existence 
to a quasi-subspace structure, similarly as discussed in \cite{univuk},  
with a matrix subblock as in (\ref{eq_2x2blocks}) but which is still very 
weakly coupled (by many indirect network links) to the core space.

%%%%%%%%%%%%%%%%%%%%%%%%%%%%%%%%%%%%%%%%%%%%%%%%%%%%%%%%%


\begin{thebibliography}{99}
\bibitem{timeorderdiscuss} This number depends on the exact time ordering 
  which is used and which is not unique because many papers are published 
  at the same time and the order between them is not specified. We have 
  chosen a time ordering where between these papers, degenerate in 
  publication time, the initial node order of the raw data is kept.
\bibitem{nonzeroentry} Note that some of the non-vanishing components 
of the iteration vector $S_0^i e$ may become very small, e.g. 
$\sim 10^{-100}$. In this context we count such components still as occupied 
despite their small size and $N_i$ is the number of nodes which can be 
reached from some arbitrary other node after $i$ iterations with the 
matrix $S_0$.
\bibitem{discuss_prefactor} 
In \cite{integers} a set of vectors without this prefactor 
was used but this provided a representation matrix which is numerically 
unstable for a direct diagonalization. The prefactor $c_{j-1}^{-1}$ 
ensures that the representation matrix is numerically (rather) stable and 
of course both matrices are mathematically related by a similarity 
transformation and have identical eigenvalues.
\end{thebibliography}
\end{document}